\documentclass[11pt,a4paper,english]{article}
\usepackage{jheppub} % use jheppub.sty
\usepackage{amsfonts,amsmath,amssymb}
\usepackage[mathscr]{euscript}
\usepackage{stmaryrd}
\usepackage{esint}

\title{Effective gravitational couplings of four-dimensional $\mathcal{N}=2$
supersymmetric gauge theories}

\author[a,b]{Jan Manschot}
\author[c]{, Gregory W. Moore}
\author[c]{and Xinyu Zhang}
\affiliation[a]{School of Mathematics, Trinity College, Dublin 2, Ireland}
\affiliation[b]{Hamilton Mathematical Institute, Trinity College, Dublin 2, Ireland}
\affiliation[c]{New High Energy Theory Center and Department of Physics and Astronomy, Rutgers University, \\ Piscataway, New Jersey 08854, USA}
\emailAdd{manschot@maths.tcd.ie}
\emailAdd{gwmoore@physics.rutgers.edu}
\emailAdd{zhangxinyuphysics@gmail.com}

\abstract{
The low energy effective couplings of a four-dimensional $\mathcal{N}=2$
supersymmetric gauge theory to topological invariants of the background
gravitational field are described by two functions $A$ and $B$.
These two functions play an important role in the study of topologically
twisted four-dimensional $\mathcal{N}=2$ supersymmetric gauge theories
and in the computation of central charges of $\mathcal{N}=2$ superconformal
theories. In this paper, we compute $A$ and $B$ from the partition
function in the $\Omega$-background for $\mathrm{SU}(2)$ gauge theories.
Our results not only confirm the predicted expressions of the effective
gravitational couplings, but also give the previously undetermined
overall multiplicative factors. We also analyze $A$ and $B$ for
the $\mathrm{SU}(N)$ super-Yang-Mills theory, and confirm all the previous predictions.
}

\begin{document}
\maketitle

%%%%%%%%%%%%%%%%%%%%Body of the paper%%%%%%%%%%%%%%%%%%%%

\section{Introduction}

The work of Seiberg and Witten \cite{Seiberg:1994rs,Seiberg:1994aj}
on four-dimensional $\mathcal{N}=2$ $\mathrm{SU}(2)$ supersymmetric
gauge theories using holomorphy and electric-magnetic duality has
revolutionized our understanding of non-perturbative dynamics in quantum
field theory. After a quarter century of hard work, the Seiberg-Witten
solution has been generalized to a large class of $\mathcal{N}=2$
theories.

The Coulomb moduli space $\mathcal{M}$, parameterized by a set of
gauge-invariant order parameters $u=\left\{ u_{1},\cdots,u_{r}\right\} $,
is a complex manifold whose dimension is the rank $r$ of the gauge
group. At a generic point in $\mathcal{M}$, the gauge group is broken
to a maximal torus $\mathrm{U}(1)^{r}$. We can choose a duality frame
with local special coordinates $a=\left\{ a_{1},\cdots,a_{r}\right\} $,
and the low energy effective theory is described in terms of $r$
abelian vector multiplets. The perturbative corrections to the low
energy effective prepotential $\mathcal{F}$ arise only at the one
loop order, while non-perturbative corrections are entirely from instantons.
It is remarkable that $\mathcal{F}$ can be solved exactly, and the
solution is elegantly encoded in the Seiberg-Witten geometry. At
singular loci $\mathcal{D}=\left\{\mathcal{D}_{s}\right\}$
in $\mathcal{M}$ extra massless particles appear.

Meanwhile, the achievement of Seiberg and Witten has also led to enormous
advances in the theory of four-manifolds. Following the earlier development
of topological field theory pioneered by Witten \cite{Witten:1988ze},
the famous Donaldson invariants of four-manifolds \cite{DonaldsonKronheimer}
can be interpreted physically as correlation functions in the topologically
twisted $\mathcal{N}=2$ $\mathrm{SU}(2)$ super-Yang-Mills theory.
With the understanding of the low energy effective dynamics of the
theory, an alternative formulation of the Donaldson invariants was
conjectured in terms of the Seiberg-Witten invariants \cite{Witten:1994cg}.
Subsequently, a physical derivation of the conjecture was given in
\cite{Moore:1997pc} and later extended and clarified in \cite{Losev:1997tp,Losev:1997bz,Marino:1998bm,Takasaki:1998vm,Marino:1998rg,Labastida:1998sk,Marino:1998eg,Marino:1998tb,Tan:2009qq,Korpas:2017qdo,Moore:2017cmm,Korpas:2018dag,Korpas:2019ava,Korpas:2019cwg}.

The path integral of the topologically twisted low energy effective
theory on a curved four-manifold $X$ receives two different contributions,
one from an integral over the Coulomb branch (often called the $u$-plane
integral), and the other from Seiberg-Witten invariants associated
to extra massless particles. Hence, the Donaldson-Witten partition
function $Z_{\mathrm{DW}}$, which is a generating function of the
Donaldson invariants, takes the form
\begin{equation}
Z_{\mathrm{DW}}=Z_{u}+\sum_{s}Z_{\mathrm{SW},s},\label{eq:ZDW}
\end{equation}
where $Z_{u}$ is the contribution from the $u$-plane, and
$Z_{\mathrm{SW},s}$ is the Seiberg-Witten contribution from the singular locus $\mathcal{D}_{s}$.
When $b_{1}(X)=0$ and $b_{2}^{+}(X)=1$, the expression of $Z_{u}$
is given by
\begin{equation}
Z_{u}=K_{u}\int\left[dad\bar{a}\right]A(u)^{\chi}B(u)^{\sigma}\Psi\left[\mathcal{K}\right].
\end{equation}
The normalization factor $K_{u}$ is chosen so that $Z_{u}$ is dimensionless.
The measure factor $A(u)^{\chi}B(u)^{\sigma}$ is holomorphic in $u$,
and encodes the couplings of the low energy effective theory to topological
invariants of the background gravitational field, where $\chi$ and
$\sigma$ are the Euler characteristic and the signature of the four-manifold,
respectively,
\begin{equation}
\chi=\frac{1}{32\pi^{2}}\int \mathrm{tr} R\wedge\tilde{R},\quad\sigma=\frac{1}{24\pi^{2}}\int \mathrm{tr} R\wedge R.
\end{equation}
The term $\Psi\left[\mathcal{K}\right]$ comes essentially from the
evaluation of the photon partition function of the low energy effective
abelian gauge theory, and takes the form of a Siegel-Narain theta
function with kernel $\mathcal{K}$ depending on the inserted observables
\cite{Korpas:2019ava,Korpas:2019cwg}.

It was found by Shapere and Tachikawa \cite{Shapere:2008zf} that
the functions $A$ and $B$ appearing in the topologically twisted
theory can be used to compute the central charges of the physical
$\mathcal{N}=2$ superconformal theory that corresponds to a superconformal
point in the Coulomb moduli space of an $\mathcal{N}=2$ supersymmetric
gauge theory. By definition, the $\mathtt{c}$ and $\mathtt{a}$ central
charges are coefficients of the Weyl tensor and the Euler density
associated with the curvature of the background gravitational field
in the conformal anomaly,
\begin{equation}
\left\langle T_{\phantom{\mu}\mu}^{\mu}\right\rangle =\frac{\mathtt{c}}{16\pi^{2}}\left(\mathrm{Weyl}\right)^{2}-\frac{\mathtt{a}}{16\pi^{2}}\left(\mathrm{Euler}\right),\label{eq:T}
\end{equation}
where the Weyl tensor and the Euler density associated with the curvature
of the background gravitational field are given by
\begin{equation}
\left(\mathrm{Weyl}\right)^{2}=R_{\mu\nu\rho\sigma}^{2}-2R_{\mu\nu}^{2}+\frac{1}{3}R^{2},\quad\left(\mathrm{Euler}\right)=R_{\mu\nu\rho\sigma}^{2}-4R_{\mu\nu}^{2}+R^{2}.
\end{equation}
We introduce a background $\mathrm{SU}(2)_{R}$ gauge connection with field strength $W_{\mu\nu}^{a}$. We can get the anomaly for the $\mathrm{U}(1)_{R}$-current
$\mathcal{R}^{\mu}$ from the conformal anomaly using the superconformal
algebra \cite{Anselmi:1997am,Anselmi:1997ys,Kuzenko:1999pi},
\begin{equation}
\partial_{\mu}\mathcal{R}^{\mu}=\frac{\mathtt{c}-\mathtt{a}}{8\pi^{2}}R_{\mu\nu\rho\sigma}\tilde{R}^{\mu\nu\rho\sigma}+\frac{2\mathtt{a}-\mathtt{c}}{8\pi^{2}}W_{\mu\nu}^{a}\tilde{W}_{a}^{\mu\nu}.\label{eq:anomalyR}
\end{equation}
We perform a topological twist by setting the $\mathrm{SU}(2)_{R}$ gauge connection equal to the self-dual part of the spin connection. Integrating the anomaly equation (\ref{eq:anomalyR}) over the four-manifold, we obtain the $\mathrm{U}(1)_{R}$
anomaly of the vacuum
\begin{equation}
\Delta\mathcal{R}=2\left(2\mathtt{a}-\mathtt{c}\right)\chi+3\mathtt{c}\sigma.\label{eq:DeltaR1}
\end{equation}
On the other hand, if there are $r$ free vector multiplets and $h$
free neutral hypermultiplets in the low energy effective theory, we
can also read the $\mathrm{U}(1)_{R}$ anomaly from the low energy
effective action on the curved manifold. The $\mathrm{U}(1)_{R}$
anomalies of a free vector multiplet and a free hypermultiplet are
$\frac{1}{2}\left(\chi+\sigma\right)$ and $\frac{1}{4}\sigma$, respectively.
If the $\mathrm{U}(1)_{R}$-charges of $A$ and $B$ are $\mathcal{R}(A)$
and $\mathcal{R}(B)$, respectively, then the $\mathrm{U}(1)_{R}$
anomaly of the vacuum is also given by
\begin{equation}
\Delta\mathcal{R}=\mathcal{R}(A)\chi+\mathcal{R}(B)\sigma+\frac{r}{2}\left(\chi+\sigma\right)+\frac{h}{4}\sigma.\label{eq:DeltaR2}
\end{equation}
Combining (\ref{eq:DeltaR1}) and (\ref{eq:DeltaR2}), we obtain the
central charges
\begin{equation}
\mathtt{a}=\frac{1}{4}\mathcal{R}(A)+\frac{1}{6}\mathcal{R}(B)+\frac{5}{24}r+\frac{1}{24}h,\quad\mathtt{c}=\frac{1}{3}\mathcal{R}(B)+\frac{1}{6}r+\frac{1}{12}h.\label{eq:ac}
\end{equation}

Our interest in the $u$-plane integral also comes from the study
of the non-trivial six-dimensional $\mathcal{N}=\left(2,0\right)$
superconformal theories, whose existence is one of the most striking
predictions of string theory \cite{Witten:1995zh,Seiberg:1996vs,Seiberg:1996qx}. We can realize the six-dimensional $\mathcal{N}=\left(2,0\right)$ theory of type
$A_{N-1}$ using a stack of $N$ parallel M5-branes \cite{Strominger:1995ac}.
After compactifying on a Riemann surface $\mathcal{C}$ with punctures,
we can obtain a four-dimensional $\mathcal{N}=2$ supersymmetric field
theory $\mathcal{T}_{N}^{\mathrm{4d}}\left[\mathcal{C}\right]$ \cite{Klemm:1996bj,Witten:1997sc,Gaiotto:2009we,Gaiotto:2009hg}. Such theories are called the $\mathcal{N}=2$ theories of class $\mathcal{S}$. In particular, for $\mathcal{N}=2$ superconformal theories of class $\mathcal{S}$, the space of coupling constants can be identified with Teichm\"{u}ller space, the universal covering space of the moduli space of complex structures of punctured Riemann surfaces. Moreover, the ultraviolet S-duality group is identified with the group of large diffeomorphisms acting on $\mathcal{C}$ that leave its complex structure fixed.

The low energy effective theory of $\mathcal{T}_{N}^{\mathrm{4d}}\left[\mathcal{C}\right]$
on the Coulomb branch is governed by a single smooth M5-brane wrapped
on the Seiberg-Witten curve $\Sigma\subset T^{*}\mathcal{C}$ \cite{Witten:1997sc},
which is an algebraic curve depending on the Coulomb branch order
parameters $u$, the masses $m_{f}$ of the hypermultiplets, as well
as the cutoff $\Lambda$ for asymptotically free theories or the ultraviolet
 coupling $\tau_{\mathrm{UV}}$ for superconformal theories. The low
energy dynamics of a single M5-brane is governed by a six-dimensional
$\mathcal{N}=\left(2,0\right)$ abelian tensor multiplet, which can
be described using an action principle \cite{Henneaux:1988gg,Perry:1996mk,Pasti:1996vs,Pasti:1997gx,Bandos:1997ui,Aganagic:1997zq,Bandos:1997gm,Belov:2006jd}.
Now we put the six-dimensional $\mathcal{N}=\left(2,0\right)$ abelian
theory $\mathcal{T}^{\mathrm{6d}}$ on $X\times\Sigma$. The R-symmetry
group of $\mathcal{T}^{\mathrm{6d}}$ is $\mathrm{Spin}(5)_{R}$,
which has a subgroup $\mathrm{Spin}(3)_{R}\times\mathrm{Spin}(2)_{R}\cong\mathrm{SU}(2)_{R}\times\mathrm{U}(1)_{R}$.
Let $\mathrm{SU}(2)_{+}^{\prime}$ and $\mathrm{U}(1)_{\Sigma}^{\prime}$
be the diagonal subgroups of $\mathrm{SU}(2)_{+}\times\mathrm{SU}(2)_{R}$
and $\mathrm{U}(1)_{\Sigma}\times\mathrm{U}(1)_{R}$, respectively.
We can apply the standard procedure of topological twisting 
\footnote{We would like to emphasize that we use Lie groups rather than Lie
algebras in the procedure of topological twisting: This procedure requires
the introduction of a bundle with connection associated with the R-symmetry
group, and together with an isomorphism of bundles such that relevant
connections are mapped to each other under the isomorphism. In the
study of the Donaldson-Witten theory, the required $\mathrm{SU}(2)_{R}$ bundle
might not exist, but the $\mathrm{SO}(3)_{R}$ bundle associated
to the adjoint representation always exists. One can choose an isomorphism
of this adjoint bundle with the bundle of self-dual two-forms. Then
one puts a connection on the adjoint $\mathrm{SO}(3)_{R}$ bundle
so that under this isomorphism we get the Levi-Civita connection on
the self-dual two-forms. In our case, however, one must choose a $\mathrm{Spin}(5)_{R}$
bundle together with a reduction of the structure group to 
$\mathrm{Spin}(3)_{R}\times\mathrm{Spin}(2)_{R}$.} 
and replace the holonomy group $\mathrm{SU}(2)_{-}\times\mathrm{SU}(2)_{+}$
of $X$ and the holonomy group $\mathrm{U}(1)_{\Sigma}$ of $\Sigma$
with $\mathrm{SU}(2)_{-}\times\mathrm{SU}(2)_{+}^{\prime}$ and $\mathrm{U}(1)_{\Sigma}^{\prime}$,
respectively. In order to compute the partition function of $\mathcal{T}^{\mathrm{6d}}$
on $X\times\Sigma$, we can either first compactify $\mathcal{T}^{\mathrm{6d}}$
on $\Sigma$ to obtain the low energy effective theory $\mathcal{T}_{\mathrm{IR}}^{\mathrm{4d}}\left[\Sigma\right]$
of the ultraviolet theory $\mathcal{T}_{N}^{\mathrm{4d}}\left[\mathcal{C}\right]$
on $X$ with Donaldson-Witten twist, or first compactify $\mathcal{T}^{\mathrm{6d}}$
on $X$ to get a two-dimensional $\mathcal{N}=(0,2)$ theory $\mathcal{T}^{\mathrm{2d}}\left[X\right]$ on $\Sigma$ with half-twist \cite{Witten:1991zz,Witten:1993yc}.
Because of the topological nature of the setup, the integrand of
the $u$-plane integral of $\mathcal{T}_{\mathrm{IR}}^{\mathrm{4d}}\left[\Sigma\right]$
on $X$ should coincide with a correlation function in $\mathcal{T}^{\mathrm{2d}}\left[X\right]$ on $\Sigma$. Therefore, we can deduce $A$ and $B$ using this correspondence
by changing the topology of $X$. The relation of four-manifold invariants 
with two-dimensional $\mathcal{N}=(0,2)$  has been discussed in \cite{MooreNidaev,Gadde:2013sca,Dedushenko:2017tdw,Nidaiev:2019mzt}. 
In spite of this work, the full derivation of the Coulomb branch integrals for topologically 
twisted class $\mathcal{S}$ theories remains to be completed. We will leave a discussion of this interesting topic for another occasion. 

Based on the requirements of holomorphy, the $\mathrm{U}(1)_{R}$
R-symmetry, and the single-valuedness of the integrand of $Z_{u}$,
the general forms of $A$ and $B$ were predicted to be \cite{Witten:1995gf,Moore:1997pc,Losev:1997tp,Marino:1998bm,Shapere:2008zf}
\begin{equation}
A=\alpha\left(\det\frac{du_{i}}{da_{j}}\right)^{\frac{1}{2}},\quad B=\beta\Delta^{\frac{1}{8}}.\label{eq:AB}
\end{equation}
Here $\Delta$ is the physical discriminant, which is a holomorphic
function with first order zeroes at the locus $\left\{ u_{s}\right\} $
where extra particles becomes massless. For $\mathrm{SU}(2)$ gauge
theories, we normalize $\Delta$ as
\begin{equation}
\Delta=\prod_{s}\left(u-u_{s}\right).
\end{equation}
The physical discriminant can be different from the mathematical discriminant
of the Seiberg-Witten curve for two reasons \cite{Shapere:2008zf}.
First, the Seiberg-Witten curve is not unique for a given $\mathcal{N}=2$
gauge theory \cite{Martinec:1995by,Schulze:1997ex,Caceres:1998bc}.
Different forms give the same solution to the low energy effective
theory and the same BPS spectrum, but may give different mathematical
discriminants. Second, it is not guaranteed that all the cycles of
the Seiberg-Witten curve correspond to physical states, and if some
zeroes of the mathematical discriminant do not indicate the appearance
of extra massless particles, we should not include them when we compute
$\Delta$.

The overall multiplicative factors $\alpha$ and $\beta$ in (\ref{eq:AB})
are constants on the Coulomb branch that have not been determined
yet. In principle, they can depend on the theory, the masses of hypermultiplets,
and also on the cutoff $\Lambda$ for asymptotically free theories
or the ultraviolet  coupling $\tau_{\mathrm{UV}}$ for superconformal
theories. For the $\mathrm{SU}(2)$ super-Yang-Mills theory, we need
to choose $\left(K_{u},\alpha,\beta\right)$ so that the partition
function (\ref{eq:ZDW}) matches precisely with known results of Donaldson
invariants from the mathematical literature. The choice made in \cite{Korpas:2019cwg}
is that 
\footnote{We have rescaled here the $\Lambda$ and $a$ of \cite{Korpas:2019cwg}
to compare their $\alpha$ and $\beta$ to ours,
\begin{equation}
\Lambda_{\mathrm{KMMN}}=\sqrt{2}\Lambda,\quad a_{\mathrm{KMMN}}=\frac{a}{\sqrt{2}},
\end{equation}
while $u_{\mathrm{KMMN}}=u$.
}

\begin{equation}
K_{u}=2^{-\frac{5}{2}}\Lambda^{-3},\quad\alpha=2^{\frac{1}{8}}e^{-\frac{\pi\mathrm{i}}{8}}\pi^{-\frac{1}{2}},\quad\beta=2^{\frac{5}{8}}e^{-\frac{\pi\mathrm{i}}{8}}\pi^{-\frac{1}{2}}.\label{eq:KabSU2}
\end{equation}
For other theories, there is no mathematical result to compare with.
It was predicted in \cite{Marino:1998bm} that the $N$-dependence
of $\alpha$ and $\beta$ in the $\mathrm{SU}(N)$ super-Yang-Mills
theory should be
\begin{equation}
\alpha(N)=e^{\kappa_{1}^{(\alpha)}N+\kappa_{2}^{(\alpha)}N^{2}},\quad\beta(N)=e^{\kappa_{1}^{(\beta)}N+\kappa_{2}^{(\beta)}N^{2}},\label{eq:alphabetaN}
\end{equation}
where $\kappa_{1}^{(\alpha,\beta)}$ and $\kappa_{2}^{(\alpha,\beta)}$
are $N$-independent constants that can depend on $\Lambda$. It was
also argued in \cite{Marino:1998tb} that $\alpha$ and $\beta$ are
independent of masses for asymptotically free theories.

Up to now, almost nothing has been known about $\alpha$ and $\beta$
for superconformal theories. It is certainly interesting to figure
out how $\alpha$ and $\beta$ depend on the parameters of the theory,
especially on the conformal manifold for superconformal theories.
It was proposed by Labastida and Lozano \cite{Labastida:1998sk} that
for the $\mathrm{SU}(2)$ $\mathcal{N}=2^{*}$ theory 
\footnote{In order to compare the $\alpha$ and $\beta$ of \cite{Labastida:1998sk}
to ours, we need to rescale the $m$, $u$ and $a$ of \cite{Labastida:1998sk} by
\begin{equation}
m_{\mathrm{LL}}=\sqrt{2}m,\quad u_{\mathrm{LL}}=2u,\quad a_{\mathrm{LL}}=2a.
\end{equation}
We should also notice that the discriminant used in \cite{Labastida:1998sk}
is the mathematical discriminant of the Seiberg-Witten curve, which
is related to the physical discriminant $\Delta$ used in this paper by
\begin{equation}
\Delta_{\mathrm{LL}}=8\eta\left(\tau_{\mathrm{UV}}\right)^{12}\Delta.
\end{equation}
}
\begin{equation}
K_{u}\alpha^{\chi}\beta^{\sigma}=-\frac{4\mathrm{i}}{\pi}2^{\frac{3}{8}\chi + \frac{21}{16}\sigma}{\mu}^{2\chi+3\sigma}\eta\left(\tau_{\mathrm{UV}}\right)^{-3\chi-\frac{3}{2}\sigma}m^{\frac{1}{8}\sigma},\label{eq:ABLL}
\end{equation}
so that the Donaldson-Witten partition function $Z_{\mathrm{DW}}$ in the massless limit coincides with the Vafa-Witten partition function
\cite{Vafa:1994tf} on $K3$ manifolds. The function ${\mu}$ was not
determined since $2\chi+3\sigma=0$ for $K3$ manifolds. Clearly, at least one of $K_{u}$, $\alpha$ and $\beta$ must depend nontrivially on $\tau_{\mathrm{UV}}$. We expect that for general superconformal theories of class $\mathcal{S}$, $\alpha$ and $\beta$ are automorphic forms on the Teichm\"{u}ller space.

Given the importance of $A$ and $B$, it is definitely beneficial
to cross-check the prediction (\ref{eq:AB}) using other approaches. In this paper, we shall specify the gravitational
background to be the $\Omega$-background of $\mathbb{R}^{4}\cong\mathbb{C}^{2}$
and apply the powerful instanton counting techniques \cite{Nekrasov:2002qd}
to compute $A$ and $B$. Our strategy is to expand the exact partition
function $\mathcal{Z}$ in the $\Omega$-background around the flat
space limit $\varepsilon_{1},\varepsilon_{2}\to0$,
\begin{equation}
\varepsilon_{1}\varepsilon_{2}\log\mathcal{Z}=-\mathcal{F}+\left(\varepsilon_{1}+\varepsilon_{2}\right)\mathcal{H}+\varepsilon_{1}\varepsilon_{2}\log A+\frac{\varepsilon_{1}^{2}+\varepsilon_{2}^{2}}{3}\log B+\cdots,\label{eq:expansion}
\end{equation}
where $\varepsilon_{1}$ and $\varepsilon_{2}$ are two deformation
parameters of the $\Omega$-background, and $\cdots$ includes higher
order terms in $\varepsilon_{1},\varepsilon_{2}$ that are irrelevant
to our problem. The leading term coincides with the low energy effective
prepotential $\mathcal{F}$ \cite{Nekrasov:2002qd}. This gives us
an opportunity to derive rigorously the Seiberg-Witten geometry for
a large class of $\mathcal{N}=2$ theories. In fact, by the saddle
point analysis, the partition function $\mathcal{Z}$ in the limit
$\varepsilon_{1},\varepsilon_{2}\to0$ is dominated by a particular
instanton configuration determined by the limit shape equations, whose
solution leads to the Seiberg-Witten curve \cite{Nekrasov:2003rj,Shadchin:2004yx,Shadchin:2005cc,Nekrasov:2012xe,Zhang:2019msw}.
A priori, we cannot rule out the next-to-leading order term $\mathcal{H}$,
but it vanishes in every example we will be dealing
with. The identification of the next two terms follows from the equivariant
Euler characteristic and the equivariant signature of $\mathbb{C}^{2}$
\cite{Nakajima:2003uh},
\begin{equation}
\chi\left(\mathbb{C}^{2}\right)=\varepsilon_{1}\varepsilon_{2},\quad\sigma\left(\mathbb{C}^{2}\right)=\frac{\varepsilon_{1}^{2}+\varepsilon_{2}^{2}}{3}.\label{eq:invariants}
\end{equation}
Hence from the partition function $\mathcal{Z}$ we can directly compute
$A$ and $B$, and determine $\alpha$ and $\beta$ from first principles.
\footnote{In fact, the information of $A$ and $B$ can be extracted from $\mathcal{Z}$ using two linearly independent limits for the $\varepsilon_{1}$ and $\varepsilon_{2}$. For example, we can use the topological string limit $\varepsilon_{+}\to 0$ and the Nekrasov-Shatashvili limit $\varepsilon_{2}\to 0$ \cite{Nekrasov:2009rc}.}
Similar expansions were performed in \cite{Billo:2013fi,Billo:2013jba},
leading to a modular anomaly equation. However, they simply disregarded
the $a$-independent terms and the important $\mathrm{U}(1)$ factors
in their analysis. These terms can be ignored when we are only interested
in the dynamics of the theory, but they are crucial to our problem.

There is an important subtlety regarding the normalization involved in our analysis. Since the
partition function $\mathcal{Z}$ is naturally normalized to have
vanishing mass dimension, we see that the mass dimensions of $A$
and $B$ are zero. On the other hand, in the standard normalization
of the $u$-plane integral, $A$ and $B$ have nonzero mass dimensions. In
order to resolve this problem, we notice that we only consider the
situation $b_{1}(X)=0$ and $b_{2}^{+}(X)=1$. Therefore, we have
$\chi+\sigma=4$, and there is a normalization ambiguity \cite{Moore:2017cmm}
\begin{equation}
\left(K_{u},\alpha,\beta\right)\sim\left(\kappa^{-4}K_{u},\kappa\alpha,\kappa\beta\right).\label{eq:ambiguity}
\end{equation}
We can use this ambiguity to relate the results computed from $\mathcal{Z}$
with those appearing in $Z_{u}$. Notice that the ratio $\beta/\alpha$ is unambiguous.

To illustrate our method, we shall mainly focus on the simple examples
of $\mathrm{SU}(2)$ gauge theories. We can write down the partition
function $\mathcal{Z}$ explicitly up to an arbitrary order of the
instanton number. We then compare our results with those computed
using the Seiberg-Witten curve. In this way, we successfully confirm
the prediction (\ref{eq:AB}), and obtain the overall factors $\alpha$
and $\beta$. We also confirm (\ref{eq:AB}) and (\ref{eq:alphabetaN}) for the $\mathrm{SU}(N)$ super-Yang-Mills
theory.

The rest of the paper is organized as follows. In section \ref{sec:Omega}
we summarize the useful results of the partition function of the four-dimensional
$\mathcal{N}=2$ supersymmetric gauge theory in the $\Omega$-background.
In section \ref{sec:SYM} we consider the $\mathrm{SU}(2)$ super-Yang-Mills
theory. In section \ref{sec:adj} we deal with the $\mathrm{SU}(2)$
theory with an adjoint hypermultiplet. In section \ref{sec:fund}
we study the $\mathrm{SU}(2)$ gauge theory with four fundamental
hypermultiplets. In section \ref{sec:Pert} we analyze $A$ and $B$ in the $\mathrm{SU}(N)$ super-Yang-Mills
theory. We conclude in section \ref{sec:Discussion} with a discussion
of some subtleties of our results as well as an outlook of future
directions. In appendix \ref{app:Gamma} we discuss the definition
and the expansion of the special function $\gamma_{\varepsilon_{1},\varepsilon_{2}}\left(x;\Lambda\right)$.
In appendix \ref{app:Period} we review how to compute the period
integrals on an elliptic curve. In appendix \ref{app:Modular} we
collect a few essential aspects of the theory of modular forms and
Jacobi theta functions. In appendix \ref{app:Weierstrass} we review
Weierstrass's elliptic function.

\section{Partition function in the $\Omega$-background \label{sec:Omega}}

Let us consider the four-dimensional $\mathcal{N}=2$ supersymmetric
gauge theory with gauge group $G=\mathrm{SU}(N)$ and massive hypermultiplets 
\footnote{In this paper, we always consider full hypermultiplets. See \cite{Hollands:2010xa,Hollands:2011zc,Zhang:2019msw} for work on half-hypermultiplets in the $\Omega$-background.}
in a representation $\mathfrak{R}$ of $G$. We can decompose $\mathfrak{R}$ into irreducible representations of $G$,
\begin{equation}
\mathfrak{R} = \bigoplus_f \mathfrak{R}_f.
\end{equation}
We require the beta function of the gauge coupling
constant $g$ to be non-positive so that we can have a well-defined
microscopic theory,
\begin{equation}
\Lambda\frac{\partial g}{\partial\Lambda}=-\frac{g^{3}}{16\pi^{2}}\left(2N-2T\left(\mathfrak{R}\right)\right)\leq0,
\end{equation}
where $\Lambda$ is the cutoff scale, and $T\left(\mathfrak{R}\right)$ is the quadratic Casimir of the representation $\mathfrak{R}$ satisfying $T\left(\mathfrak{R}_1\oplus \mathfrak{R}_2\right)=T\left(\mathfrak{R_1}\right)+T\left(\mathfrak{R}_2\right)$. In this paper, we are mainly interested in the adjoint and fundamental representations,
\begin{equation}
T\left(\mathrm{adj}\right)=N,\quad T\left(\mathrm{fund}\right)=\frac{1}{2}.
\end{equation}
For asymptotically free theories, we define the instanton counting
parameter $\mathtt{q}$ to be
\begin{equation}
\mathtt{q}=\Lambda^{2N-2 T\left(\mathfrak{R}\right)}.
\end{equation}
For superconformal theories we have the ultraviolet complexified coupling
\begin{equation}
\tau_{\mathrm{UV}}=\frac{\vartheta_{\mathrm{UV}}}{2\pi}+\frac{4\pi\mathrm{i}}{g_{\mathrm{UV}}^{2}},
\end{equation}
where $\vartheta_{\mathrm{UV}}$ and $g_{\mathrm{UV}}$ are the ultraviolet
theta angle and gauge coupling constant, respectively, and we define
\begin{equation}
\mathtt{q}=e^{2\pi\mathrm{i}\tau_{\mathrm{UV}}}.
\end{equation}
We choose the vacuum expectation value of the scalar field $\phi$
in the vector multiplet to be the local special coordinates $a$ on
the Coulomb branch,
\begin{equation}
a=\left\langle \phi\right\rangle =\mathrm{diag}\left(a_{1},\cdots,a_{N}\right).
\end{equation}

It is useful to introduce the $\Omega$-deformation of the theory
\cite{Moore:1997dj,Nekrasov:2002qd}, so that the Poincar\'{e} symmetry
of $\mathbb{R}^{4} \cong \mathbb{C}^{2}$ is broken in a rotationally covariant way, while
still preserving a particular linear combination of supercharges
\begin{equation}
\mathcal{Q}=\bar{Q}+\Omega^{\mu}Q_{\mu}.
\end{equation}
Here $\bar{Q}$ and $Q_{\mu}$ are the scalar and vector supercharges
in the topologically twisted $\mathcal{N}=2$ theories \cite{Witten:1988ze},
and $\Omega^{\mu}\partial_{\mu}$ is the Killing vector generating
the $\mathrm{U}(1)^{2}$ isometry of $\mathbb{C}^{2}$,
\begin{equation}
\Omega^{\mu}\partial_{\mu}=\mathrm{i}\varepsilon_{1}\left(z_{1}\frac{\partial}{\partial z_{1}}-\bar{z}_{1}\frac{\partial}{\partial\bar{z}_{1}}\right)+\mathrm{i}\varepsilon_{2}\left(z_{2}\frac{\partial}{\partial z_{2}}-\bar{z}_{2}\frac{\partial}{\partial\bar{z}_{2}}\right).
\end{equation}
The supersymmetric action in the $\Omega$-background can be constructed
from the flat space action by replacing $\phi$ by an operator \cite{Nekrasov:2010ka}
\begin{equation}
\phi\mapsto\phi+\Omega^{\mu}D_{\mu}.
\end{equation}
We also define
\begin{equation}
\varepsilon_{\pm}=\frac{\varepsilon_{1}\pm\varepsilon_{2}}{2}.
\end{equation}
Clearly, the $\Omega$-background is closely related to the topological
twist, since $\mathcal{Q}$ will become the usual scalar supercharge
$\bar{Q}$ used in the topologically twisted theory when we take the
limit $\varepsilon_{1},\varepsilon_{2}\to0$.

Using the powerful localization techniques, the partition function
$\mathcal{Z}$ in the $\Omega$-background can be calculated exactly
and is given by a product of the classical, one-loop and instanton
contributions \cite{Nekrasov:2002qd},
\begin{equation}
\mathcal{Z}=\mathcal{Z}^{\mathrm{cl}}\mathcal{Z}^{\mathrm{1-loop}}\mathcal{Z}^{\mathrm{inst}}.
\end{equation}
It is convenient to start with the gauge group $\mathrm{U}(N)$. The
classical contribution is given by
\begin{equation}
\mathcal{Z}^{\mathrm{cl}}\left(a,\mathtt{q};\varepsilon_{1},\varepsilon_{2}\right)=\mathtt{q}^{-\frac{1}{2\varepsilon_{1}\varepsilon_{2}}\sum_{i=1}^{N}a_{i}^{2}}.\label{eq:Zcl}
\end{equation}
The one-loop contributions of the vector multiplet and the hypermultiplet
in the fundamental or adjoint representation are given by
\begin{eqnarray}
\mathcal{Z}^{\mathrm{1-loop,vec}} & = & \prod_{i<j}\exp\left[-\gamma_{\varepsilon_{1},\varepsilon_{2}}\left(a_{i}-a_{j};\Lambda\right)-\gamma_{\varepsilon_{1},\varepsilon_{2}}\left(a_{i}-a_{j}-2\varepsilon_{+};\Lambda\right)\right],\\
\mathcal{Z}^{\mathrm{1-loop,fund}} & = & \prod_{i=1}^{N}\exp\left[\gamma_{\varepsilon_{1},\varepsilon_{2}}\left(a_{i}+m-\varepsilon_{+};\Lambda\right)\right],\label{eq:1loopfund}\\
\mathcal{Z}^{\mathrm{1-loop,adj}} & = & \prod_{i,j=1}^{N}\exp\left[\gamma_{\varepsilon_{1},\varepsilon_{2}}\left(a_{i}-a_{j}+m-\varepsilon_{+};\Lambda\right)\right],\label{eq:1loopadj}
\end{eqnarray}
where the definition and basic properties of the special function
$\gamma_{\varepsilon_{1},\varepsilon_{2}}\left(x;\Lambda\right)$
are given in appendix \ref{app:Gamma}. For asymptotically free theories,
it is convenient to absorb the classical contribution into the one-loop
contribution by redefining $\Lambda$. Notice that we do not include
the contributions from $i=j$ for the vector multiplet but we should
include them for the adjoint hypermultiplet.

The instanton partition function is given by
\begin{equation}
\mathcal{Z}^{\mathrm{inst}}=\sum_{k=0}^{\infty}\mathtt{q}^{k}\int_{\mathfrak{M}_{k}}\mathrm{e}\left(\mathscr{E}_{\mathrm{matter}}\to\mathfrak{M}_{k}\right),
\end{equation}
where $\mathfrak{M}_{k}$ is the moduli space of framed noncommutative
$\mathrm{U}(N)$ instantons on $\mathbb{C}^{2}$ with instanton charge
$k$, 
\footnote{Here the noncommutative deformation is introduced to resolve the singularities
of the moduli space due to point-like instantons.} 
and $\mathrm{e}\left(\mathscr{E}_{\mathrm{matter}}\to\mathfrak{M}_{k}\right)$
is the equivariant Euler class of the matter bundle whose fiber is
the space of the virtual zero modes for the Dirac operator associated
with the hypermultiplet in the instanton background. $\mathcal{Z}^{\mathrm{inst}}$
can be evaluated using the equivariant localization formula. The fixed
points in $\mathfrak{M}=\cup_{k}\mathfrak{M}_{k}$ are labeled by
$N$-tuple of Young diagrams $\vec{Y}=\left\{ Y^{(1)},\cdots Y^{(N)}\right\} $,
and the equivariant Euler class at fixed points can be computed from
the equivariant Chern characters. Then $\mathcal{Z}^{\mathrm{inst}}$
is reduced to a statistical sum over Young diagrams,
\begin{equation}
\mathcal{Z}^{\mathrm{inst}}=\sum_{\vec{Y}}\mathtt{q}^{\left|\vec{Y}\right|}z_{\mathrm{vec}}\left(a,\vec{Y}\right)\prod_{f}z_{\mathrm{hyper}}^{\mathfrak{R}_{f}}\left(a,m_{f},\vec{Y}\right),
\end{equation}
where $\left|\vec{Y}\right|$ is the total number of boxes in $N$
Young diagrams. We introduce the conversion operator $\epsilon$ which
maps characters into weights,
\begin{equation}
\epsilon\left\{ \sum_{i}n_{i}e^{x_{i}}\right\} =\prod_{i}x_{i}^{n_{i}},
\end{equation}
and the dual operator $\vee$,
\begin{equation}
\left(\sum_{i}n_{i}e^{x_{i}}\right)^{\vee}=\left(\sum_{i}n_{i}e^{-x_{i}}\right).
\end{equation}
The contributions from the vector multiplet and the hypermultiplet
in the fundamental or adjoint representation can be written compactly
as \cite{Nekrasov:2002qd,Nekrasov:2012xe,Nekrasov:2015wsu}
\begin{eqnarray}
z_{\mathrm{vec}}\left(a,\vec{Y}\right) & = & \epsilon\left\{ -\mathscr{N}\mathscr{K}_{\vec{Y}}^{\vee}-e^{2\varepsilon_{+}}\mathscr{K}_{\vec{Y}}\mathscr{N}^{\vee}+\mathscr{P}\mathscr{K}_{\vec{Y}}\mathscr{K}_{\vec{Y}}^{\vee}\right\} ,\\
z_{\mathrm{hyper}}^{\mathrm{fund}}\left(a,m,\vec{Y}\right) & = & \epsilon\left\{ e^{m+\varepsilon_{+}}\mathscr{K}_{\vec{Y}}\right\} ,\label{eq:instfund}\\
z_{\mathrm{hyper}}^{\mathrm{adj}}\left(a,m,\vec{Y}\right) & = & \epsilon\left\{ e^{m-\varepsilon_{+}}\left(\mathscr{N}\mathscr{K}_{\vec{Y}}^{\vee}+e^{2\varepsilon_{+}}\mathscr{K}_{\vec{Y}}\mathscr{N}^{\vee}-\mathscr{P}\mathscr{K}_{\vec{Y}}\mathscr{K}_{\vec{Y}}^{\vee}\right)\right\} ,\label{eq:instadj}
\end{eqnarray}
where
\begin{eqnarray}
\mathscr{N} & = & \sum_{i=1}^{N}e^{a_{i}},\\
\mathscr{K}_{\vec{Y}} & = & \sum_{i=1}^{N}\sum_{(x,y)\in Y^{(i)}}e^{a_{i}+\varepsilon_{1}\left(x-1\right)+\varepsilon_{2}\left(y-1\right)},\\
\mathscr{P} & = & \left(1-e^{\varepsilon_{1}}\right)\left(1-e^{\varepsilon_{2}}\right).
\end{eqnarray}
This expression of $\mathcal{Z}^{\mathrm{inst}}$ can reproduce the standard expression in terms of arm and leg lengths using combinatorial formulas \cite{nakajima1999lectures,Nakajima:2003pg}.

It is worth emphasizing that the masses $m_{f}$ appearing in (\ref{eq:1loopfund})(\ref{eq:1loopadj})(\ref{eq:instfund})(\ref{eq:instadj})
differ from the masses $m_{f}^{\prime}$ in the original paper \cite{Nekrasov:2002qd,Nekrasov:2003rj}
by a constant shift of $\varepsilon_{+}$ \cite{Gottsche:2010ig,Okuda:2010ke,Huang:2011qx},
\begin{equation}
m_{f}=m_{f}^{\prime}+\varepsilon_{+}.\label{eq:mshift}
\end{equation}
This shift is due to the fact that the scalars in a hypermultiplet become spinors in the Donaldson-Witten twist, and the Dirac complex is the Dolbeault complex twisted by the square-root of the canonical bundle of the four-manifold. This shift
can often be ignored in many applications of the $\Omega$-background,
since it will not modify the dynamics of the theory on flat space
where $\varepsilon_{+}=0$. However, the functions $A$ and $B$ are defined in the Donaldson-Witten twist, and it is necessary to use $m_{f}$ as the mass parameters.

When we move from the gauge group $\mathrm{U}(N)$ to $\mathrm{SU}(N)$,
we have to modify carefully the partition function in the $\Omega$-background.
First of all, we need to set
\begin{equation}
\sum_{i=1}^{N}a_{i}=0.
\end{equation}
In particular, for $G=\mathrm{SU}(2)$, we take
\begin{equation}
\left\langle \phi\right\rangle =\begin{pmatrix}a & 0\\
0 & -a
\end{pmatrix}.
\end{equation}
Second, while the tensor product $\mathrm{fund}\otimes\overline{\mathrm{fund}}$
of the fundamental and the anti-fundamental representations gives
the adjoint representation for the group $\mathrm{U}(N)$, we have
to subtract the trivial representation to get the adjoint representation
for the group $\mathrm{SU}(N)$. Therefore, the one-loop contribution
of the $\mathrm{SU}(N)$ adjoint hypermultiplet is given by (\ref{eq:1loopadj})
divided by $\exp\left[\gamma_{\varepsilon_{1},\varepsilon_{2}}\left(m-\varepsilon_{+};\Lambda\right)\right]$.
Finally, we need to factor out the residual contribution of the $\mathrm{U}(1)\subset\mathrm{U}(N)$
gauge field from the instanton partition function,
\begin{equation} 
\mathcal{Z}_{\mathrm{U}(N)}^{\mathrm{inst}}=\mathcal{Z}_{\mathrm{SU}(N)}^{\mathrm{inst}}\mathcal{Z}_{\mathrm{extra}}^{\mathrm{inst}}.
\end{equation}
The explicit expression of $\mathcal{Z}_{\mathrm{extra}}^{\mathrm{inst}}$
was first proposed in \cite{Alday:2009aq}, and later derived from
the non-perturbative Dyson-Schwinger equations \cite{Nekrasov:2015wsu,Jeong:2017mfh,Nekrasov:2017gzb}.
For the $\mathrm{SU}(2)$ gauge theory with one adjoint hypermultiplet
of mass $m$, we have
\begin{equation}
\mathcal{Z}_{\mathrm{extra}}^{\mathrm{inst}}=\left[\prod_{n=1}^{\infty}\left(1-\mathtt{q}^{n}\right)\right]^{-\frac{2}{\varepsilon_{1}\varepsilon_{2}}\left(m+\varepsilon_{-}\right)\left(m-\varepsilon_{-}\right)},
\end{equation}
and for the $\mathrm{SU}(2)$ gauge theory with four fundamental hypermultiplets
of masses $m_{1},m_{2},m_{3},m_{4}$, we have 
\footnote{We consider here four fundamental hypermultiplets rather than two
fundamental and two anti-fundamental hypermultiplets as in \cite{Alday:2009aq}.
The factor $\mathcal{Z}_{\mathrm{extra}}^{\mathrm{inst}}$ breaks
the $\mathrm{Spin}(8)$ symmetry of the masses. However, the breaking
only affects the low energy effective prepotential $\mathcal{F}$
in the expansion (\ref{eq:expansion}), and leads to a constant shift
in $\mathcal{F}$.}
\begin{equation}
\mathcal{Z}_{\mathrm{extra}}^{\mathrm{inst}}=\left(1-\mathtt{q}\right)^{\frac{2}{\varepsilon_{1}\varepsilon_{2}}\left(\frac{m_{1}+m_{2}}{2}+\varepsilon_{+}\right)\left(\frac{m_{3}+m_{4}}{2}+\varepsilon_{+}\right)}.
\end{equation}

In the following three sections, we will focus on $\mathrm{SU}(2)$ gauge theories. For this gauge group, we can also take the advantage of the equivalence $\mathrm{SU}(2)\cong \mathrm{Sp}(1)$ and directly perform the computation using the $\mathrm{Sp}(1)$ gauge theory \cite{Marino:2004cn,Nekrasov:2004vw,Shadchin:2004yx,Fucito:2004gi}. It is known that the $\mathrm{Sp}(1)$ instanton moduli space looks rather different from the $\mathrm{SU}(2)$ instanton moduli space \cite{Atiyah:1978ri}. Nevertheless, it was shown in \cite{Hollands:2010xa} that these two partition functions agree, possibly up to an $a$-independent factor and after a nontrivial mapping of parameters. Therefore, if (\ref{eq:AB}) can be demonstrated using the $\mathrm{SU}(2)$ partition function, it automatically holds if we use the $\mathrm{Sp}(1)$ partition function. We should also not worry about the $a$-independent factor, since $\alpha$ and $\beta$ depend on the precise microscopic definition of the theory. The choice made in this paper is a natural choice of ultraviolet regularizations, and it turns out to be consistent with all the previous results.

\section{The $\mathrm{SU}(2)$ super-Yang-Mills theory \label{sec:SYM}}

The simplest but most important example is the $\mathrm{SU}(2)$ super-Yang-Mills theory.

\subsection{Expansion of the partition function}

The partition function of the theory in the $\Omega$-background is given in section \ref{sec:Omega} with $\mathcal{Z}_{\mathrm{extra}}^{\mathrm{inst}}=1$.

It is straightforward to compute the expansion (\ref{eq:expansion}).
The leading term gives the low energy effective prepotential $\mathcal{F}$,
\begin{equation}
\mathcal{F}=-4a^{2}\left(\log\left(\frac{2a}{\Lambda}\right)-\frac{3}{2}\right)+\frac{\Lambda^{4}}{2a^{2}}+\frac{5\Lambda^{8}}{64a^{6}}+\frac{3\Lambda^{12}}{64a^{10}}+\frac{1469\Lambda^{16}}{32768a^{14}}+\mathcal{O}\left(\frac{\Lambda^{20}}{a^{18}}\right),\label{eq:Fpure}
\end{equation}
from which we can compute the Coulomb moduli order parameter $u$ \cite{Matone:1995rx,Flume:2004rp},
\begin{eqnarray}
u & = & \frac{1}{2}\left\langle \mathrm{Tr}\phi^{2}\right\rangle =\frac{1}{4}\Lambda\frac{\partial\mathcal{F}}{\partial\Lambda}\nonumber \\
 & = & a^{2}+\frac{\Lambda^{4}}{2a^{2}}+\frac{5\Lambda^{8}}{32a^{6}}+\frac{9\Lambda^{12}}{64a^{10}}+\frac{1469\Lambda^{16}}{8192a^{14}}+\mathcal{O}\left(\frac{\Lambda^{20}}{a^{18}}\right).\label{eq:uapure}
\end{eqnarray}
The next-to-leading order term $\mathcal{H}=0$. In fact, the perturbative
contribution vanishes because of the expansion (\ref{eq:gammaexp})
of the function $\gamma_{\varepsilon_{1},\varepsilon_{2}}\left(x;\Lambda\right)$,
and the instanton contribution also vanishes since $\mathcal{Z}^{\mathrm{inst}}$
is invariant under $\left(\varepsilon_{1},\varepsilon_{2}\right)\to\left(-\varepsilon_{1},-\varepsilon_{2}\right)$.
The second order terms in the expansion (\ref{eq:expansion}) are
\begin{align}
\log A & =\frac{1}{2}\log\left(\frac{2a}{\Lambda}\right)-\frac{\Lambda^{4}}{4a^{4}}-\frac{19\Lambda^{8}}{64a^{8}}-\frac{47\Lambda^{12}}{96a^{12}}-\frac{15151\Lambda^{16}}{16384a^{16}}+\mathcal{O}\left(\frac{\Lambda^{20}}{a^{20}}\right),\label{eq:logApure}\\
\log B & =\frac{1}{2}\log\left(\frac{2a}{\Lambda}\right)-\frac{3\Lambda^{4}}{8a^{4}}-\frac{63\Lambda^{8}}{128a^{8}}-\frac{55\Lambda^{12}}{64a^{12}}-\frac{55335\Lambda^{16}}{32768a^{16}}+\mathcal{O}\left(\frac{\Lambda^{20}}{a^{20}}\right).\label{eq:logBpure}
\end{align}

\subsection{Comparison to the prediction}

In order to compare our results computed from the partition function
in the $\Omega$-background to the prediction (\ref{eq:AB}), we consider
the Seiberg-Witten curve
\begin{equation}
\label{SWcurve}
y^{2}=\left(x^{2}-u\right)^{2}-4\Lambda^{4},
\end{equation}
with the Seiberg-Witten differential $\lambda$ determined by
\begin{equation}
\frac{\partial\lambda}{\partial u}=\frac{1}{2\pi\mathtt{i}}\frac{dx}{y}.
\end{equation}
Using the result of the period integral (\ref{eq:PA1}), we have
\begin{eqnarray}
\label{daduofa}
\frac{da}{du} & = & \frac{1}{2\pi\mathtt{i}}\oint_{A}\frac{dx}{y}\nonumber \\
 & = & \left(\sqrt{u-2\Lambda^{2}}+\sqrt{u+2\Lambda^{2}}\right)^{-1}{_{2}}F_{1}\left(\frac{1}{2},\frac{1}{2},1,\left(\frac{\sqrt{u-2\Lambda^{2}}-\sqrt{u+2\Lambda^{2}}}{\sqrt{u-2\Lambda^{2}}+\sqrt{u+2\Lambda^{2}}}\right)^{2}\right)\nonumber \\
 & = & \frac{1}{2\sqrt{u}}+\frac{3\Lambda^{4}}{8u^{5/2}}+\frac{105\Lambda^{8}}{128u^{9/2}}+\frac{1155\Lambda^{12}}{512u^{13/2}}+\frac{225225\Lambda^{16}}{32768u^{17/2}}+\mathcal{O}\left(\frac{\Lambda^{20}}{u^{21/2}}\right).
\end{eqnarray}
Using techniques from the theory of elliptic curves, one can express the observable
$da/du$ as a function of the complex structure $\tau$ of the curve
(\ref{SWcurve}) in closed form, \footnote{This equation and (\ref{uoftau}) appear different from those in \cite{Korpas:2019cwg} due to the different normalizations of $a$, $\Lambda$ and $\tau$.}
\begin{equation}
\label{daduoftau}
\Lambda\frac{da}{du}=\frac{1}{4}\theta_2(\tau)^2,
\end{equation}
where $\theta_2$ is one of the Jacobi theta functions defined in (\ref{Jthetas}).

Integrating with respect to $u$, we get
\begin{equation}
a(u)=\sqrt{u}-\frac{\Lambda^{4}}{4u^{3/2}}-\frac{15\Lambda^{8}}{64u^{7/2}}-\frac{105\Lambda^{12}}{256u^{11/2}}-\frac{15015\Lambda^{16}}{16384u^{15/2}}+\mathcal{O}\left(\frac{\Lambda^{20}}{u^{19/2}}\right),
\end{equation}
and its inverse function is
\begin{equation}
\label{uofa}
u(a)=a^{2}+\frac{\Lambda^{4}}{2a^{2}}+\frac{5\Lambda^{8}}{32a^{6}}+\frac{9\Lambda^{12}}{64a^{10}}+\frac{1469\Lambda^{16}}{8192a^{14}}+\mathcal{O}\left(\frac{\Lambda^{20}}{a^{18}}\right).
\end{equation}
As a function of $\tau$, $u$ reads
\begin{equation}
\label{uoftau}
\frac{u(\tau)}{\Lambda^2}=4\frac{\theta_3(\tau)^4}{\theta_2(\tau)^4}-2.
\end{equation}

Returning to the results for the instanton partition function
in the $\Omega$-background, we recognize that the expansion in (\ref{uofa})
matches with the result (\ref{eq:uapure}). From (\ref{daduofa}), we determine
\begin{equation}
\log\left(\frac{du}{da}\right)=\log(2a)-\frac{\Lambda^{4}}{2a^{4}}-\frac{19\Lambda^{8}}{32a^{8}}-\frac{47\Lambda^{12}}{48a^{12}}-\frac{15151\Lambda^{16}}{8192a^{16}}+\mathcal{O}\left(\frac{\Lambda^{20}}{a^{20}}\right).\label{eq:ASWpure}
\end{equation}
On the other hand, the singularities of the Coulomb branch are at
$u=\pm2\Lambda^{2}$ where we have extra massless BPS states. Therefore
the physical discriminant is given by $\Delta=u^{2}-4\Lambda^{4}$,
whose logarithm is given by
\begin{equation}
\log\Delta=4\log(a)-\frac{3\Lambda^{4}}{a^{4}}-\frac{63\Lambda^{8}}{16a^{8}}-\frac{55\Lambda^{12}}{8a^{12}}-\frac{55335\Lambda^{16}}{4096a^{16}}+\mathcal{O}\left(\frac{\Lambda^{20}}{a^{20}}\right).\label{eq:BSWpure}
\end{equation}

By comparing (\ref{eq:ASWpure}) and (\ref{eq:BSWpure}) with $A$
and $B$ given in (\ref{eq:logApure}) and (\ref{eq:logBpure}), respectively,
we find
\begin{equation}
A=\Lambda^{-\frac{1}{2}}\left(\frac{du}{da}\right)^{\frac{1}{2}},\quad B=\sqrt{2}\Lambda^{-\frac{1}{2}}\Delta^{\frac{1}{8}},\label{eq:ABpure}
\end{equation}
which reproduce (\ref{eq:AB}). We also match the unambiguous ratio
with (\ref{eq:KabSU2}),
\begin{equation}
\frac{\beta}{\alpha}=\sqrt{2}=\frac{2^{\frac{5}{8}}e^{-\frac{\pi\mathrm{i}}{8}}\pi^{-\frac{1}{2}}}{2^{\frac{1}{8}}e^{-\frac{\pi\mathrm{i}}{8}}\pi^{-\frac{1}{2}}}.
\end{equation}
Finally, we note that we can express $\tau=\frac{1}{4\pi
  i}\frac{\partial^2\mathcal{F}}{\partial a^2}$ as an expansion in
$\Lambda/a$ using (\ref{eq:Fpure}). Substitution of this expansion in
(\ref{daduoftau}) and (\ref{uoftau}) reproduces the expansions in (\ref{daduofa})
and (\ref{uoftau}).

\section{The $\mathrm{SU}(2)$ $\mathcal{N}=2^{*}$ theory \label{sec:adj}}

The simplest $\mathcal{N}=2$ superconformal theory is the $\mathcal{N}=4$
super-Yang-Mills theory, which is the $\mathcal{N}=2$ gauge theory
with one adjoint hypermultiplet. We turn on the $\mathcal{N}=2$ invariant
bare mass term and the resulting theory is often called the $\mathcal{N}=2^{*}$
theory. In this section, we take the gauge group $G=\mathrm{SU}(2)$,
and denote the mass by $m$. In the class $\mathcal{S}$ construction,
this theory arises by compactifying the six-dimensional $(2,0)$ theory
of type $A_{1}$ on a torus with one puncture.

\subsection{Expansion of the partition function}

We can compute the expansion (\ref{eq:expansion}) of the partition
function in the $\Omega$-background. The leading term is the low
energy effective prepotential $\mathcal{F}$. Up to $\mathcal{O}\left(\mathtt{q}^{5}\right)$,
it is given explicitly by
\begin{eqnarray}
\label{2*Fofa}
\mathcal{F} & = & a^{2}\log\mathtt{q}+m^{2}\left(\log\frac{2a}{\Lambda}+\frac{1}{2}\log\frac{m}{\Lambda}-\frac{3}{4}\right)\nonumber \\
 &  & +\frac{m^{4}}{a^{2}}\left(-\frac{1}{48}+\frac{1}{2}\mathtt{q}+\frac{3}{2}\mathtt{q}^{2}+2\mathtt{q}^{3}+\frac{7}{2}\mathtt{q}^{4}+\mathcal{O}\left(\mathtt{q}^{5}\right)\right)\nonumber \\
 &  & +\frac{m^{6}}{a^{4}}\left(-\frac{1}{960}-\frac{3}{4}\mathtt{q}^{2}-4\mathtt{q}^{3}-\frac{45}{4}\mathtt{q}^{4}+\mathcal{O}\left(\mathtt{q}^{5}\right)\right)\nonumber \\
 &  & +\frac{m^{8}}{a^{6}}\left(-\frac{1}{10752}+\frac{5}{64}\mathtt{q}^{2}+\frac{5}{2}\mathtt{q}^{3}+\frac{1095}{64}\mathtt{q}^{4}+\mathcal{O}\left(\mathtt{q}^{5}\right)\right)+\mathcal{O}\left(\frac{m^{10}}{a^{8}}\right),
\end{eqnarray}
where we organize $\mathcal{F}$ as a series in inverse powers of
$a^{2}$. Because of the S-duality of the ultraviolet theory, we expect
that the $\mathtt{q}$-series in each parentheses is the first few
terms of a quasi-modular form, which can be written in terms of the
Eisenstein series $E_{2}$, $E_{4}$ and $E_{6}$. Indeed, we can
complete the $\mathtt{q}$-series to get
\begin{eqnarray}
\mathcal{F} & = & a^{2}\log\mathtt{q}+m^{2}\left(\log\frac{2a}{\Lambda}+\frac{1}{2}\log\frac{m}{\Lambda}-\frac{3}{4}\right)-\frac{m^{4}E_{2}}{48a^{2}}\nonumber \\
 &  & -\frac{m^{6}\left(5E_{2}^{2}+E_{4}\right)}{5760a^{4}}-\frac{m^{8}\left(175E_{2}^{3}+84E_{2}E_{4}+11E_{6}\right)}{2903040a^{6}}+\mathcal{O}\left(\frac{m^{10}}{a^{8}}\right).\label{eq:Fadj}
\end{eqnarray}
The appearance of the quasi-modular form $E_{2}$ is unavoidable in
order for $\mathcal{F}$ to transform properly under S-duality \cite{Minahan:1997if}.
The $\Lambda$ dependent part of $\mathcal{F}$ is
\begin{equation}
\mathcal{F}\sim-\frac{3}{2}m^{2}\log\Lambda. \label{FLambda}
\end{equation}
If we weakly gauge the $\mathrm{U}(1)$ flavor symmetry, then $m$
can be viewed as the vacuum expectation value of the corresponding
vector multiplet. The hypermultiplet transforms in the adjoint representation
of the gauge group $\mathrm{SU}(2)$. From the $\mathcal{N}=2$ preserving superpotential
\begin{equation}
W = \sqrt{2} \mathrm{Tr} \tilde{Q} \Phi Q + m \mathrm{Tr} \tilde{Q} Q ,
\end{equation}
we know that the hypermultiplet has charge $\pm 1$ under
this $\mathrm{U}(1)$. We can get the coefficient of the one-loop
beta function for the $\mathrm{U}(1)$ coupling constant from
\begin{equation}
\Lambda\frac{\partial^{3}\mathcal{F}}{\partial\Lambda\partial m^{2}}=-3,
\end{equation}
where the sign is opposite to that of an asymptotically free theory,
and $3$ is the dimension of the adjoint representation of $\mathrm{SU}(2)$.

Using the derivatives of the Eisenstein series (\ref{eq:dE}), we
obtain the Coulomb branch order parameter
\begin{eqnarray}
u & = & \frac{1}{2}\left\langle \mathrm{Tr}\phi^{2}\right\rangle =\mathtt{q}\frac{\partial\mathcal{F}}{\partial\mathtt{q}}\nonumber \\
 & = & a^{2}+\frac{m^{4}\left(-E_{2}^{2}+E_{4}\right)}{576a^{2}}+\frac{m^{6}\left(-5E_{2}^{3}+3E_{2}E_{4}+2E_{6}\right)}{34560a^{4}}\nonumber \\
 &  & +\frac{m^{8}\left(-35E_{2}^{4}+7E_{2}^{2}E_{4}+10E_{4}^{2}+18E_{2}E_{6}\right)}{2322432a^{6}}+\mathcal{O}\left(\frac{m^{10}}{a^{8}}\right),\label{eq:uadj}
\end{eqnarray}
which is independent of $\Lambda$.

We then go beyond the leading order in the expansion (\ref{eq:expansion}).
It is interesting that we still have $\mathcal{H}=0$ in the presence
of the adjoint hypermultiplet. At the second order, we have two terms
which are our main interest,
\begin{eqnarray}
\log A & = & \frac{1}{2}\log\frac{2a}{\Lambda}+\frac{m^{4}}{a^{4}}\left(-\frac{1}{4}\mathtt{q}-\frac{3}{2}\mathtt{q}^{2}-3\mathtt{q}^{3}-7\mathtt{q}^{4}+\mathcal{O}\left(\mathtt{q}^{5}\right)\right)\nonumber \\
 &  & +\frac{m^{6}}{a^{6}}\left(\frac{3}{2}\mathtt{q}^{2}+12\mathtt{q}^{3}+45\mathtt{q}^{4}+\mathcal{O}\left(\mathtt{q}^{5}\right)\right)\nonumber \\
 &  & +\frac{m^{8}}{a^{8}}\left(-\frac{19}{64}\mathtt{q}^{2}-12\mathtt{q}^{3}-\frac{3405}{32}\mathtt{q}^{4}+\mathcal{O}\left(\mathtt{q}^{5}\right)\right)+\mathcal{O}\left(\frac{m^{10}}{a^{10}}\right),
\end{eqnarray}
and
\begin{eqnarray}
\log B & = & \frac{3}{4}\log\frac{2a}{\Lambda}+\frac{1}{8}\log\frac{m}{\Lambda}+\frac{m^{2}}{a^{2}}\left(-\frac{1}{32}+\frac{3}{4}\mathtt{q}+\frac{9}{4}\mathtt{q}^{2}+3\mathtt{q}^{3}+\frac{21}{4}\mathtt{q}^{4}+\mathcal{O}\left(\mathtt{q}^{5}\right)\right)\nonumber \\
 &  & +\frac{m^{4}}{a^{4}}\left(-\frac{1}{256}-\frac{3}{8}\mathtt{q}-\frac{81}{16}\mathtt{q}^{2}-\frac{39}{2}\mathtt{q}^{3}-\frac{843}{16}\mathtt{q}^{4}+\mathcal{O}\left(\mathtt{q}^{5}\right)\right)\nonumber \\
 &  & +\frac{m^{6}}{a^{6}}\left(-\frac{1}{1536}+\frac{195}{64}\mathtt{q}^{2}+\frac{75}{2}\mathtt{q}^{3}+\frac{12465}{64}\mathtt{q}^{4}+\mathcal{O}\left(\mathtt{q}^{5}\right)\right)\nonumber \\
 &  & +\frac{m^{8}}{a^{8}}\left(-\frac{1}{8192}-\frac{63}{128}\mathtt{q}^{2}-\frac{441}{16}\mathtt{q}^{3}-\frac{83097}{256}\mathtt{q}^{4}+\mathcal{O}\left(\mathtt{q}^{5}\right)\right)+\mathcal{O}\left(\frac{m^{10}}{a^{10}}\right).
\end{eqnarray}
Similar to the treatment of $\mathcal{F}$, we complete each $\mathtt{q}$-series
into a quasi-modular form,
\begin{eqnarray}
\log A & = & \frac{1}{2}\log\frac{2a}{\Lambda}+\frac{m^{4}\left(E_{2}^{2}-E_{4}\right)}{1152a^{4}}+\frac{m^{6}\left(5E_{2}^{3}-3E_{2}E_{4}-2E_{6}\right)}{34560a^{6}}\nonumber \\
 &  & +\frac{m^{8}\left(203E_{2}^{4}-28E_{2}^{2}E_{4}-67E_{4}^{2}-108E_{2}E_{6}\right)}{9289728a^{8}}+\mathcal{O}\left(\frac{m^{10}}{a^{10}}\right),\label{eq:Aadj}\\
\log B & = & \frac{3}{4}\log\frac{2a}{\Lambda}+\frac{1}{8}\log\frac{m}{\Lambda}-\frac{m^{2}E_{2}}{32a^{2}}-\frac{m^{4}\left(E_{2}^{2}+E_{4}\right)}{512a^{4}}-\frac{m^{6}\left(25E_{2}^{3}+48E_{2}E_{4}+17E_{6}\right)}{138240a^{6}}\nonumber \\
 &  & -\frac{m^{8}\left(1225E_{2}^{4}+3332E_{2}^{2}E_{4}+1055E_{4}^{2}+1948E_{2}E_{6}\right)}{61931520a^{8}}+\mathcal{O}\left(\frac{m^{10}}{a^{10}}\right).\label{eq:Badj}
\end{eqnarray}

We can get the pure $\mathcal{N}=2$ super-Yang-Mills theory from the $\mathcal{N}=2^{*}$
theory by taking a certain decoupling limit. This limit is not manifest
in (\ref{eq:Fadj}) since it is written in the limit $m/a\to0$. The
expression of $\mathcal{F}$ in the limit $m/a\to\infty$ is given
by
\begin{eqnarray}
\mathcal{F} & = & m^{2}\left(\frac{3}{2}\log\left(\frac{m}{\Lambda}\right)-\frac{9}{4}\right)+a^{2}\log\frac{\mathtt{q}m^{4}}{\Lambda^{4}}-4a^{2}\left(\log\left(\frac{2a}{\Lambda}\right)-\frac{3}{2}\right)\nonumber \\
 &  & +\mathtt{q}\frac{m^{4}}{2a^{2}}+\mathtt{q}^{2}\left(\frac{5m^{8}}{64a^{6}}-\frac{3m^{6}}{4a^{4}}+\frac{3m^{4}}{2a^{2}}\right)+\mathcal{O}\left(\mathtt{q}^{3},\frac{a}{m}\right).\label{eq:Flargem}
\end{eqnarray}
Therefore, if we take the limit
\begin{equation}
m\to\infty,\quad\mathtt{q}\to 0,\quad\mathtt{q}m^{4}=\Lambda^{4},\label{eq:decouple}
\end{equation}
the effective prepotential (\ref{eq:Flargem}) becomes (\ref{eq:Fpure}) up to a constant,
\begin{equation}
\mathcal{F}_{\mathcal{N}=2^{*}} \to \mathcal{F}_{\mathrm{SYM}} + m^{2}\left(\frac{3}{2}\log\left(\frac{m}{\Lambda}\right)-\frac{9}{4}\right),\label{eq:urelation}
\end{equation}
and the relation between the order parameters $u_{\mathcal{N}=2^{*}}$
and $u_{\mathrm{SYM}}$ is given by
\begin{equation}
u_{\mathcal{N}=2^{*}} \to u_{\mathrm{SYM}} - \frac{3 m^2}{8}.\label{eq:urelation}
\end{equation}

Similarly, we can consider the limit (\ref{eq:decouple}) for $\log A$ and $\log B$,
\begin{eqnarray}
\log A & = & \frac{1}{2}\log\left(\frac{2a}{\Lambda}\right)-\mathtt{q}\frac{m^{4}}{4a^{4}}-\mathtt{q}^{2}\left(\frac{19m^{8}}{64a^{8}}-\frac{3m^{6}}{2a^{6}}+\frac{3m^{4}}{2a^{4}}\right)+\mathcal{O}\left(\mathtt{q}^{3},\frac{a}{m}\right)\nonumber \\
 & \to & \frac{1}{2}\log\left(\frac{2a}{\Lambda}\right)-\frac{\Lambda^{4}}{4a^{4}}-\frac{19\Lambda^{8}}{64a^{8}}+\mathcal{O}\left(\frac{\Lambda^{12}}{a^{12}}\right),\\
\log B & = & \frac{1}{2}\log\left(\frac{2a}{\Lambda}\right)+\frac{3}{8}\log\left(\frac{m}{\Lambda}\right)-\mathtt{q}\left(\frac{3m^{4}}{8a^{4}}-\frac{3m^{2}}{4a^{2}}\right)\nonumber \\
 &  & -\mathtt{q}^{2}\left(\frac{63m^{8}}{128a^{8}}-\frac{195m^{6}}{64a^{6}}+\frac{81m^{4}}{16a^{4}}-\frac{9m^{2}}{4a^{2}}\right)+\mathcal{O}\left(\mathtt{q}^{3},\frac{a}{m}\right)\nonumber \\
 & \to & \frac{1}{2}\log\left(\frac{2a}{\Lambda}\right)+\frac{3}{8}\log\left(\frac{m}{\Lambda}\right)-\frac{3\Lambda^{4}}{8a^{4}}-\frac{63\Lambda^{8}}{128a^{8}}+\mathcal{O}\left(\frac{\Lambda^{12}}{a^{12}}\right).
\end{eqnarray}
Hence we have
\begin{equation}
A_{\mathcal{N}=2^{*}}\to A_{\mathrm{SYM}},\quad B_{\mathcal{N}=2^{*}}\to\left(\frac{m}{\Lambda}\right)^{\frac{3}{8}}B_{\mathrm{SYM}}.
\end{equation}

\subsection{Comparison to the prediction}

In order to compare our results with the prediction (\ref{eq:AB}),
we take the Seiberg-Witten curve and the Seiberg-Witten differential
to be \cite{Witten:1997sc,Gaiotto:2009we,Gaiotto:2009hg}
\begin{equation}
t^{2}=\tilde{u}-\nu m^{2}\wp\left(z;\tau_{\mathrm{UV}}\right),\quad\lambda=tdz,\label{eq:SWadj}
\end{equation}
where the parameter $\tilde{u}$ in the curve is the same as $u$
up to an additive constant,
\begin{equation}
\tilde{u}=u+m^{2}h\left(\tau_{\mathrm{UV}}\right),
\end{equation}
and $\wp\left(z;\tau_{\mathrm{UV}}\right)$ is Weierstrass's elliptic
function (see appendix \ref{app:Weierstrass} for its basic properties).
We see from the curve (\ref{eq:SWadj}) that $\tilde{u}$ is modular
under the ultraviolet S-duality transformation. The adjustable numerical
constant $\nu$ depends on the normalization and will be fixed later.
In fact, (\ref{eq:SWadj}) is a special example of the Seiberg-Witten
geometry constructed using the elliptic Calogero-Moser integrable
system \cite{DHoker:1997hut,DHoker:1998xad}. 
%By introducing $x=\wp\left(z;\tau_{\mathrm{UV}}\right)$ and $y=\wp^{\prime}\left(z;\tau_{\mathrm{UV}}\right)$, we can rewrite the Seiberg-Witten curve (\ref{eq:SWadj}) as
%\begin{equation} 
%y^{2}=4x^{3}-g_{2}x-g_{3}=4\left(x-e_{1}\right)\left(x-e_{2}\right)\left(x-e_{3}\right),
%\end{equation}
%and the Seiberg-Witten differential 
%\begin{equation}
%\lambda^{2}=\left(\tilde{u}-\nu m^{2}x\right)\left(\frac{dx}{y}\right)^{2}.
%\end{equation}

We can extract $a\left(\tilde{u}\right)$ in the usual way from the
period integral,
\begin{eqnarray}
a\left(\tilde{u}\right) & = & \frac{1}{\pi}\oint_{A}\sqrt{\tilde{u}-\nu m^{2}\wp\left(z;\tau_{\mathrm{UV}}\right)}dz\nonumber \\
 & = & \sqrt{\tilde{u}}\left(1-\frac{\nu m^{2}}{2\tilde{u}}\mathcal{P}_{1}-\frac{\nu^{2}m^{4}}{8\tilde{u}^{2}}\mathcal{P}_{2}-\frac{\nu^{3}m^{6}}{16\tilde{u}^{3}}\mathcal{P}_{3}-\frac{5\nu^{4}m^{8}}{128\tilde{u}^{4}}\mathcal{P}_{4}+\mathcal{O}\left(\frac{m^{10}}{\tilde{u}^{5}}\right)\right),\label{eq:a_adj}
\end{eqnarray}
where we define
\begin{equation}
\mathcal{P}_{n}=\frac{1}{\pi}\oint_{A}\wp^{n}\left(z;\tau_{\mathrm{UV}}\right)dz,
\end{equation}
whose explicit expressions are given in appendix \ref{app:Weierstrass}.
We can solve $\tilde{u}$ in terms of $a$ by inverting (\ref{eq:a_adj}),
\begin{eqnarray}
\tilde{u} & = & a^{2}+\nu m^{2}\mathcal{P}_{1}-\frac{\nu^{2}m^{4}\left(\mathcal{P}_{1}^{2}-\mathcal{P}_{2}\right)}{4a^{2}}+\frac{\nu^{3}m^{6}\left(2\mathcal{P}_{1}^{3}-3\mathcal{P}_{1}\mathcal{P}_{2}+\mathcal{P}_{3}\right)}{8a^{4}}\nonumber \\
 &  & -\frac{5\nu^{4}m^{8}\left(4\mathcal{P}_{1}^{4}-8\mathcal{P}_{1}^{2}\mathcal{P}_{2}+\mathcal{P}_{2}^{2}+4\mathcal{P}_{1}\mathcal{P}_{3}-\mathcal{P}_{4}\right)}{64a^{6}}+\mathcal{O}\left(\frac{m^{10}}{a^{8}}\right)\nonumber \\
 & = & a^{2}-\frac{\nu m^{2}E_{2}}{3}+\frac{\nu^{2}m^{4}\left(-E_{2}^{2}+E_{4}\right)}{36a^{2}}+\frac{\nu^{3}m^{6}\left(-5E_{2}^{3}+3E_{2}E_{4}+2E_{6}\right)}{540a^{4}}\nonumber \\
 &  & +\frac{\nu^{4}m^{8}\left(-35E_{2}^{4}+7E_{2}^{2}E_{4}+10E_{4}^{2}+18E_{2}E_{6}\right)}{9072a^{6}}+\mathcal{O}\left(\frac{m^{10}}{a^{8}}\right),\label{eq:uaadjSW}
\end{eqnarray}
which matches (\ref{eq:uadj}) if we choose
\begin{equation}
\tilde{u}=u-\frac{m^{2}E_{2}}{12},\quad\nu=\frac{1}{4}.
\end{equation}
In fact, one can give a closed form expression for $\tilde{u}$ as function of $\tau$ and $\tau_{\mathrm{UV}}$ \cite{Huang:2011qx},
\begin{equation}
\label{uoftautauUV}
\tilde{u}(\tau,\tau_{\mathrm{UV}})= -\frac{m^2}{4}
\frac{e_1(\tau_{\mathrm{UV}})^2(e_2(\tau)-e_3(\tau))+\mathrm{cyclic}}{e_1(\tau_{\mathrm{UV}})(e_2(\tau)-e_3(\tau))+\mathrm{cyclic}},
\end{equation}
where the $e_j$ are defined in (\ref{curveroots}). Note $\tilde{u}$ is
a modular form of weight 0 in $\tau$ and weight 2 in $\tau_{\mathrm{UV}}$.

Using (\ref{eq:uaadjSW}), we can compute
\begin{eqnarray}
\log\left(\frac{du}{da}\right) & = & \log\left(\frac{d\tilde{u}}{da}\right)\nonumber \\
 & = & \log2a+\frac{m^{4}\left(E_{2}^{2}-E_{4}\right)}{576a^{4}}+\frac{m^{6}\left(5E_{2}^{3}-3E_{2}E_{4}-2E_{6}\right)}{17280a^{6}}\nonumber \\
 &  & +\frac{m^{8}\left(203E_{2}^{4}-28E_{2}^{2}E_{4}-67E_{4}^{2}-108E_{2}E_{6}\right)}{4644864a^{8}}+\mathcal{O}\left(\frac{m^{10}}{a^{10}}\right).\label{eq:ASWadj}
\end{eqnarray}
From the relation (\ref{eq:urelation}), we know that in the limit (\ref{eq:decouple}),
\begin{equation}
\left(\frac{du}{da}\right)_{\mathcal{N}=2^{*}}\to\left(\frac{du}{da}\right)_{\mathrm{SYM}}.
\end{equation}
As function of $\tau$ and $\tau_\mathrm{UV}$, $da/du$ can be
expressed as
\begin{equation}
\label{2*dadu}
\frac{da}{du}=\frac{1}{4\,m\,\eta(\tau_\mathrm{UV})^6} \left( \theta_4(\tau)^4\,\theta_3(\tau_\mathrm{UV})^4-\theta_3(\tau)^4\,\theta_4(\tau_\mathrm{UV})^4\right)^{\frac{1}{2}},
\end{equation}
where $\eta$ is the Dedekind eta function given in (\ref{Deta}).

There are three singularities on the Coulomb branch where we have
extra massless particles. From (\ref{eq:SWadj}) we know that the singularities are at points
\begin{equation}
\tilde{u}=\frac{m^{2}}{4}e_{i},\quad i=1,2,3,
\end{equation}
where $e_{i}$ are defined in (\ref{eq:wp}). Therefore, the physical discriminant $\Delta$ is given by
\begin{eqnarray}
\Delta & = & \prod_{i=1}^{3}\left(\tilde{u}-\frac{m^{2}}{4}e_{i}\right)\nonumber \\
 & = & \tilde{u}^{3}-\frac{m^{2}}{4}\left(e_{1}+e_{2}+e_{3}\right)\tilde{u}^{2}+\frac{m^{4}}{16}\left(e_{1}e_{2}+e_{1}e_{3}+e_{2}e_{3}\right)\tilde{u}-\frac{m^{6}}{64}e_{1}e_{2}e_{3}\nonumber \\
 & = & \tilde{u}^{3}-\frac{m^{4}E_{4}}{48}\tilde{u}-\frac{m^{6}E_{6}}{864},\label{eq:disadj}
\end{eqnarray}
which using (\ref{uoftautauUV}) can be written as
\begin{equation}
\Delta=(2m)^6\,\eta(\tau_\mathrm{UV})^{24} \frac{\eta(\tau)^{12}}{\left(\theta_4(\tau)^4 \theta_3(\tau_\mathrm{UV})^4-\theta_3(\tau)^4 \theta_4(\tau_\mathrm{UV})^4\right)^{3}}.
\end{equation}

Substituting (\ref{eq:uaadjSW}) into (\ref{eq:disadj}), we obtain
\begin{eqnarray}
\log\Delta & = & 6\log a-\frac{m^{2}E_{2}}{4a^{2}}-\frac{m^{4}\left(E_{2}^{2}+E_{4}\right)}{64a^{4}}-\frac{m^{6}\left(25E_{2}^{3}+48E_{2}E_{4}+17E_{6}\right)}{17280a^{6}}\nonumber \\
 &  & -\frac{m^{8}\left(1225E_{2}^{4}+3332E_{2}^{2}E_{4}+1055E_{4}^{2}+1948E_{2}E_{6}\right)}{7741440a^{8}}+\mathcal{O}\left(\frac{m^{10}}{a^{10}}\right).\label{eq:BSWadj}
\end{eqnarray}
In the decoupling limit (\ref{eq:decouple}),
\begin{equation}
\label{2*Delta}
\Delta_{\mathcal{N}=2^{*}}\to\left(u-\frac{m^{2}}{4}\right)\left(u-2\Lambda^{2}\right)\left(u+2\Lambda^{2}\right) \to -\frac{m^{2}}{4}\Delta_{\mathrm{SYM}}.
\end{equation}

Comparing (\ref{eq:ASWadj})(\ref{eq:BSWadj}) with (\ref{eq:Aadj})(\ref{eq:Badj}),
we find that
\begin{equation}
A=\Lambda^{-\frac{1}{2}}\left(\frac{du}{da}\right)^{\frac{1}{2}},\quad B=2^{\frac{3}{4}}m^{\frac{1}{8}}\Lambda^{-\frac{7}{8}}\Delta^{\frac{1}{8}}.\label{eq:ABadj}
\end{equation}
We can get the unambiguous ratio
\begin{equation}
\frac{\beta}{\alpha}=2^{\frac{3}{4}}m^{\frac{1}{8}}\Lambda^{-\frac{3}{8}}. \label{ratio1}
\end{equation}
Similar to the $\Lambda$ dependence of the prepotential (\ref{FLambda}), the strange $\Lambda$ dependence of $\beta / \alpha$ can be understood as a remnant of gravitational couplings of the weakly gauged $\mathrm{U}(1)$ flavor symmetry. In fact, we can see from (\ref{eq:gammaexp}) that the remnant contribution of the adjoint hypermultiplet to $\alpha^{\chi} \beta^{\sigma}$ is $\Lambda^{-\frac{3}{8}}$, which precisely gives the $\Lambda$ dependence in $\beta / \alpha$.

On the other hand, from (\ref{eq:ABLL}) we have
\begin{equation}
\frac{\beta}{\alpha}=2^{\frac{15}{16}}\mu\eta\left(\tau_{\mathrm{UV}}\right)^{\frac{3}{2}}m^{\frac{1}{8}}. \label{ratio2}
\end{equation}
Combining (\ref{ratio1}) and (\ref{ratio2}), we get
\begin{equation}
\mu=2^{-\frac{3}{16}}\Lambda^{-\frac{3}{8}}\eta\left(\tau_{\mathrm{UV}}\right)^{-\frac{3}{2}}.
\end{equation}
Therefore, 
\begin{equation}
K_{u}\alpha^{\chi}\beta^{\sigma}=-\frac{4\mathrm{i}}{\pi} 2^{\frac{3}{4}\sigma} \Lambda^{-\frac{3}{4}\chi-\frac{9}{8}\sigma} \eta\left(\tau_{\mathrm{UV}}\right)^{-6\chi-6\sigma}m^{\frac{1}{8}\sigma}.
\end{equation}

Finally, we can express $\tau=\frac{1}{4\pi
  i}\frac{\partial^2\mathcal{F}}{\partial a^2}$ as a series in $a$
using (\ref{2*Fofa}). Substitution of this series in the closed
expressions (\ref{uoftautauUV}), (\ref{2*dadu}) and (\ref{2*Delta})
matches with the expansions (\ref{eq:uaadjSW}),  (\ref{eq:ASWadj}) and
(\ref{eq:BSWadj}).

\subsection{Mass parameter}

As stressed in section \ref{sec:Omega}, we need to be very careful
about the masses (\ref{eq:mshift}). We would like to show explicitly
in this theory that we need the mass $m$ rather than $m^{\prime}$
to get the sensible result from the point of view of the $u$-plane
integral.

It is interesting to notice that the instanton partition function
$\mathcal{Z}^{\mathrm{inst}}$ is $a$-independent if we take either
the limit $m\to0$ or $m^{\prime}\to0$. In fact, this is what we
expect, since in the massless limit we recover the $\mathcal{N}=4$
super-Yang-Mills theory, and there is no instanton corrections when
we study the dynamics of the theory.

However, if we identify $m^{\prime}$ rather than $m$ as the mass
of the hypermultiplet, we can again naively compute the expansion
(\ref{eq:expansion}). The leading term will not change, and is still
given by the prepotential. The next-to-leading order term $\mathcal{H}$
no longer vanishes,
\begin{eqnarray}
\mathcal{H} & = & -m^{\prime}\log\frac{2a}{\Lambda}-\frac{1}{2}m^{\prime}\log\frac{m^{\prime}}{\Lambda}+\frac{m^{\prime}}{2}\nonumber \\
 &  & +\frac{m^{\prime 3}}{a^{2}}\left(\frac{1}{24}-\mathtt{q}-3\mathtt{q}^{2}-4\mathtt{q}^{3}+\mathcal{O}\left(\mathtt{q}^{4}\right)\right)+\mathcal{O}\left(\frac{m^{\prime 5}}{a^{4}}\right).
\end{eqnarray}
Moreover, we have
\begin{eqnarray}
\log A & = & -\frac{1}{8}\log\left(\frac{m^{\prime2}}{\Lambda^{2}}\right)+\frac{m^{\prime2}}{a^{2}}\left(\frac{1}{16}-\frac{3}{2}\mathtt{q}-\frac{9}{2}\mathtt{q}^{2}-6\mathtt{q}^{3}+\mathcal{O}\left(\mathtt{q}^{4}\right)\right)\nonumber \\
 &  & +\frac{m^{\prime4}}{a^{4}}\left(\frac{1}{128}-\frac{1}{4}\mathtt{q}+\frac{33}{8}\mathtt{q}^{2}+27\mathtt{q}^{3}+\mathcal{O}\left(\mathtt{q}^{4}\right)\right)+\mathcal{O}\left(\frac{m^{\prime 6}}{a^{6}}\right),
\end{eqnarray}
and
\begin{eqnarray}
\log B & = & -\frac{1}{8}\log\left(\frac{m^{\prime2}}{\Lambda^{2}}\right)+\frac{m^{\prime2}}{a^{2}}\left(\frac{1}{16}-\frac{3}{2}\mathtt{q}-\frac{9}{2}\mathtt{q}^{2}-6\mathtt{q}^{3}+\mathcal{O}\left(\mathtt{q}^{4}\right)\right)\nonumber \\
 &  & +\frac{m^{\prime2}}{a^{4}}\left(\frac{1}{128}-\frac{3}{8}\mathtt{q}+\frac{27}{8}\mathtt{q}^{2}+\frac{51}{2}\mathtt{q}^{3}+\mathcal{O}\left(\mathtt{q}^{4}\right)\right)+\mathcal{O}\left(\frac{m^{\prime 6}}{a^{6}}\right).
\end{eqnarray}
Clearly, $A$ and $B$ will violate the general forms (\ref{eq:AB}).

On the other hand, we can compute the $\mathrm{U}(N)$ partition function
of the $\mathcal{N}=2^{*}$ theory with the limit $m^{\prime}\to0$
on a compact manifold, and the partition function is known to be the
generating function of the Euler characteristic of the moduli space
of unframed semi-stable equivariant torsion-free sheaves \cite{Vafa:1994tf}.
An explicit example of the $\mathrm{U}(2)$ partition function on
$\mathbb{CP}^{2}$ was given in \cite{Bershtein:2015xfa,Bershtein:2016mxz},
and the result was given in terms of mock modular forms.

For $K3$ manifolds we have $2\chi+3\sigma=0$. If we express $\chi$
and $\sigma$ of the $\Omega$-background of $\mathbb{C}^{2}$ in
terms of $\varepsilon_{1}$ and $\varepsilon_{2}$ using (\ref{eq:invariants}),
we get
\begin{equation}
0=2\chi+3\sigma=2\left(\varepsilon_{1}\varepsilon_{2}\right)+3\left(\frac{\varepsilon_{1}^{2}+\varepsilon_{2}^{2}}{3}\right)=\left(2\varepsilon_{+}\right)^{2}.
\end{equation}
Hence, $m^{\prime}=m$ and we no longer need to distinguish between
the Donaldson-Witten twist and the Vafa-Witten twist. This is the
reason why (\ref{eq:ABLL}) can make sense.

\section{The $\mathrm{SU}(2)$ theory with fundamental hypermultiplets \label{sec:fund}}

Now we consider the $\mathrm{SU}(2)$ gauge theory with $N_{f}\leq4$
fundamental hypermultiplets. We will mainly focus on the $N_{f}=4$
case which is superconformal, and we turn on mass deformations with
four masses $m_{1},m_{2},m_{3},m_{4}$. In the class $\mathcal{S}$
construction \cite{Gaiotto:2009we,Gaiotto:2009hg}, the $\mathrm{SU}(2)$
gauge theory with four fundamental hypermultiplets arises by compactifying
the six-dimensional $(2,0)$ theory of type $A_{1}$ on a sphere with
four punctures. There are three cusps in the moduli space where we have weakly coupled descriptions of the theory. For each cusp we can define a cross ratio of the four punctures. This cross ratio is identified with the instanton counting parameter $\mathtt{q}=e^{2\pi\mathrm{i}\tau_{\mathrm{UV}}}$ for that weak-coupling description.

\subsection{Expansion of the partition function}

Similar to the previous cases, we can compute the expansion (\ref{eq:expansion}).
Here we introduce the shorthand notation $\left\llbracket \ \right\rrbracket $
to indicate the sum of all terms that make $m_{1},m_{2},m_{3},m_{4}$
totally symmetric. For example,
\begin{equation}
\left\llbracket m_{1}^{2}\right\rrbracket =\sum_{i=1}^{4}m_{i}^{2},\quad\left\llbracket m_{1}^{2}m_{2}^{2}\right\rrbracket =\sum_{i<j}m_{i}^{2}m_{j}^{2}.
\end{equation}
We also define
\begin{equation}
\mathrm{Pf}m=m_{1}m_{2}m_{3}m_{4}.
\end{equation}
The $\mathrm{Spin}(8)$ flavor symmetry is broken by the masses down to a Weyl group of the $\mathrm{Spin}(8)$ symmetry, and the above combinations
of masses are Weyl-group invariant.
The explicit expression of the low energy effective prepotential $\mathcal{F}$
is given by
\begin{eqnarray}
\mathcal{F} & = & a^{2}\left(\log\frac{\mathtt{q}}{16}+\frac{\mathtt{q}}{2}+\frac{13\mathtt{q}^{2}}{64}+\frac{23\mathtt{q}^{3}}{192}\right)+\left\llbracket m_{1}^{2}\right\rrbracket \log\left(\frac{a}{\Lambda}\right)\nonumber \\
 &  & +\left(\left(m_{1}m_{2}+m_{3}m_{4}\right)\left(\frac{1}{2}\mathtt{q}+\frac{1}{4}\mathtt{q}^{2}+\frac{1}{6}\mathtt{q}^{3}\right)+\frac{1}{64}\left\llbracket m_{1}^{2}\right\rrbracket \left(\mathtt{q}^{2}+\mathtt{q}^{3}\right)\right)\nonumber \\
 &  & +\frac{1}{a^{2}}\left(-\frac{1}{12}\left\llbracket m_{1}^{4}\right\rrbracket +\frac{1}{64}\left\llbracket m_{1}^{2}m_{2}^{2}\right\rrbracket \left(\mathtt{q}^{2}+\mathtt{q}^{3}\right)\right.\nonumber \\
 &  & \left.+\mathrm{Pf}m\left(\frac{1}{2}\mathtt{q}+\frac{1}{4}\mathtt{q}^{2}+\frac{11}{64}\mathtt{q}^{3}\right)\right)+\mathcal{O}\left(\mathtt{q}^{4},\frac{m_i^6}{a^{4}}\right).
\end{eqnarray}
Note that the Weyl group symmetry is broken by the $a$-independent
expression in the second line above. This is not surprising since we 
broke the symmetry by moving to a weak-coupling cusp. \footnote{In the $\mathrm{Sp}(1)$ gauge theory description, the Weyl group of the $\mathrm{Spin}(8)$ symmetry is preserved in the weak-coupling limit. This does not lead to a contradiction because the $a$-independent part of $\mathcal{F}$ has no effect on the low energy effective action and is therefore not physical.}
Similar to the previous example, it is interesting to analyze the
$\Lambda$ dependence of $\mathcal{F}$
\begin{equation}
\mathcal{F}\sim-\left\llbracket m_{1}^{2}\right\rrbracket \log\Lambda.
\end{equation}
Now we should weakly gauge the $\mathrm{Spin}(8)$ flavor symmetry
group, which has a subgroup $\mathrm{U}(1)^{4}$. We view $m_{i}$
as the vacuum expectation value of the $i$th $\mathrm{U}(1)$ vector
multiplet. The hypermultiplet transforms in the fundamental representation
of the gauge group $\mathrm{SU}(2)$ and has charge $\pm1$ under
this $\mathrm{U}(1)$. We can get the coefficient of the one-loop
beta function for the $i$th $\mathrm{U}(1)$ coupling constant from
\begin{equation}
\Lambda\frac{\partial^{3}\mathcal{F}}{\partial\Lambda\partial m_{i}^{2}}=-2.
\end{equation}
Again the sign is opposite to that of an asymptotically free theory,
and $2$ is the dimension of the fundamental representation of $\mathrm{SU}(2)$.

The Coulomb branch order parameter $u$ is
\begin{eqnarray}
u & = & \frac{1}{2}\left\langle \mathrm{Tr}\phi^{2}\right\rangle =\mathtt{q}\frac{\partial\mathcal{F}}{\partial\mathtt{q}}\nonumber \\
 & = & a^{2}\left(1+\frac{\mathtt{q}}{2}+\frac{13\mathtt{q}^{2}}{32}+\frac{23\mathtt{q}^{3}}{64}\right)\nonumber \\
 &  & +\frac{1}{2}\left(m_{1}m_{2}+m_{3}m_{4}\right)\left(\mathtt{q}+\mathtt{q}^{2}+\mathtt{q}^{3}\right)+\left\llbracket m_{1}^{2}\right\rrbracket \left(\frac{1}{32}\mathtt{q}^{2}+\frac{3}{64}\mathtt{q}^{3}\right)\nonumber \\
 &  & +\frac{1}{a^{2}}\left(\left\llbracket m_{1}^{2}m_{2}^{2}\right\rrbracket \left(\frac{1}{32}\mathtt{q}^{2}+\frac{3}{64}\mathtt{q}^{3}\right)\right.\nonumber \\
 &  & \left.+\mathrm{Pf}m\left(\frac{1}{2}\mathtt{q}+\frac{1}{2}\mathtt{q}^{2}+\frac{33}{64}\mathtt{q}^{3}\right)\right)+\mathcal{O}\left(\mathtt{q}^{4},\frac{m_i^6}{a^{4}}\right).\label{eq:ufund}
\end{eqnarray}
Notice that this definition of $u$ breaks the Weyl group symmetry acting on the masses. We can define a new Coulomb branch order parameter $u^{\prime}$ which is invariant under the Weyl group symmetry by subtracting an $a$-independent constant from $u$,
\begin{equation}
u^{\prime} = u - \frac{\mathtt{q}\left(m_{1}m_{2}+m_{3}m_{4}\right)}{2(1-\mathtt{q})}.
\end{equation}

In this example, the vanishing of $\mathcal{H}$ is a little nontrivial. It is crucial that we factor out the residual contribution $\mathcal{Z}_{\mathrm{extra}}^{\mathrm{inst}}$. 

We have two interesting terms at the second order,
\begin{eqnarray}
\log A & = & \frac{1}{2}\log\left(\frac{2a}{\Lambda}\right)+\frac{\mathtt{q}}{4}+\frac{9\mathtt{q}^{2}}{64}+\frac{19\mathtt{q}^{3}}{192}\nonumber \\
 &  & -\frac{1}{a^{4}}\left(\frac{1}{64}\left\llbracket m_{1}^{2}m_{2}^{2}\right\rrbracket \left(\mathtt{q}^{2}+\mathtt{q}^{3}\right)\right.\nonumber \\
 &  & \left.+\mathrm{Pf}m\left(\frac{1}{4}\mathtt{q}+\frac{1}{8}\mathtt{q}^{2}+\frac{3}{32}\mathtt{q}^{3}\right)\right)+\mathcal{O}\left(\mathtt{q}^{4},\frac{m_i^6}{a^{6}}\right),\label{eq:Afund}
\end{eqnarray}
and
\begin{eqnarray}
\log B & = & \frac{3}{2}\log\left(\frac{a}{\Lambda}\right)+\frac{1}{2}\log2+\frac{3\mathtt{q}}{8}+\frac{27\mathtt{q}^{2}}{128}+\frac{19\mathtt{q}^{3}}{128}\nonumber \\
 &  & +\frac{1}{a^{2}}\left(\left\llbracket m_{1}^{2}\right\rrbracket \left(-\frac{1}{8}+\frac{3}{256}\mathtt{q}^{2}+\frac{3}{256}\mathtt{q}^{3}\right)\right)\nonumber \\
 &  & +\frac{1}{a^{4}}\left(-\frac{3}{64}\left\llbracket m_{1}^{2}m_{2}^{2}\right\rrbracket \left(\mathtt{q}^{2}+\mathtt{q}^{3}\right)\right.\nonumber \\
 &  & \left.-\mathrm{Pf}m\left(\frac{1}{16}+\frac{3}{8}\mathtt{q}+\frac{3}{16}\mathtt{q}^{2}+\frac{3}{16}\mathtt{q}^{3}\right)\right)+\mathcal{O}\left(\mathtt{q}^{4},\frac{m_i^6}{a^{6}}\right).\label{eq:Bfund}
\end{eqnarray}

\subsection{Comparison to the prediction}

Now we would like to compare our explicit results of $A$ and $B$
with the prediction (\ref{eq:AB}).

If all the hypermultiplets are massless, the Seiberg-Witten curve
is given by \cite{Seiberg:1994aj}
\begin{equation}
y^{2}=\prod_{i=1}^{3}\left(x-e_{i}\left(\tau_{\mathrm{SW}}\right)\hat{u}\right),\label{eq:SWfund0}
\end{equation}
which describes the double cover of a sphere with four punctures.
The argument $\tau_{\mathrm{SW}}$ of $e_{i}$ coincides with the
complex structure of the curve, and is the same as the low energy
effective coupling $\tau_{\mathrm{eff}}$. It takes value in the upper
half plane which is the universal cover of the punctured sphere parameterized
by $\mathtt{q}$. The coupling $\tau_{\mathrm{SW}}$ is related to the coupling $\tau_{\mathrm{UV}}$ by
\cite{Grimm:2007tm},
\begin{equation}
e^{2\pi\mathrm{i}\tau_{\mathrm{UV}}}=\frac{\theta_{2}\left(\tau_{\mathrm{SW}}\right)^{4}}{\theta_{3}\left(\tau_{\mathrm{SW}}\right)^{4}}=16q_{\mathrm{SW}}^{\frac{1}{2}}-128q_{\mathrm{SW}}+704q_{\mathrm{SW}}^{\frac{3}{2}}-3072q_{\mathrm{SW}}^{2}+11488q_{\mathrm{SW}}^{\frac{5}{2}}+\mathcal{O}\left(q_{\mathrm{SW}}^{3}\right),\label{eq:qtau}
\end{equation}
where $q_{\mathrm{SW}}=\exp\left(2\pi\mathrm{i}\tau_{\mathrm{SW}}\right)$.

When we turn on masses, the curve proposed by Seiberg and Witten \cite{Seiberg:1994aj}
is
\begin{eqnarray}
y^{2} & = & W_{1}W_{2}W_{3}+A\left(W_{1}T_{1}\left(e_{2}-e_{3}\right)+W_{2}T_{2}\left(e_{3}-e_{1}\right)+W_{3}T_{3}\left(e_{1}-e_{2}\right)\right)-A^{2}N,\nonumber \\
W_{i} & = & x-e_{i}\hat{u}-e_{i}^{2}R,\nonumber \\
A & = & \left(e_{1}-e_{2}\right)\left(e_{2}-e_{3}\right)\left(e_{3}-e_{1}\right),\nonumber \\
R & = & \frac{1}{2}\left\llbracket \hat{m}_{1}^{2}\right\rrbracket ,\nonumber \\
T_{1} & = & \frac{1}{12}\left\llbracket \hat{m}_{1}^{2}\hat{m}_{2}^{2}\right\rrbracket -\frac{1}{24}\left\llbracket \hat{m}_{1}^{4}\right\rrbracket ,\nonumber \\
T_{2,3} & = & \pm\frac{1}{2}\mathrm{Pf}\hat{m}-\frac{1}{24}\left\llbracket \hat{m}_{1}^{2}\hat{m}_{2}^{2}\right\rrbracket +\frac{1}{48}\left\llbracket \hat{m}_{1}^{4}\right\rrbracket ,\nonumber \\
N & = & \frac{3}{16}\left\llbracket \hat{m}_{1}^{2}\hat{m}_{2}^{2}\hat{m}_{3}^{2}\right\rrbracket -\frac{1}{96}\left\llbracket \hat{m}_{1}^{4}\hat{m}_{2}^{2}\right\rrbracket +\frac{1}{96}\left\llbracket \hat{m}_{1}^{6}\right\rrbracket .\label{eq:SWfund1}
\end{eqnarray}
Here the argument of $e_{i}$ is still $\tau_{\mathrm{SW}}$, but
it is no longer the complex structure of the curve, and therefore
is different from the low energy effective coupling $\tau_{\mathrm{eff}}$.
In principle, we can compare our results computed from the partition
function $\mathcal{Z}$ with the curve (\ref{eq:SWfund1}). However,
it turns out that the parameter $\hat{u}$ and the masses $\hat{m}_{i}$
in the curve (\ref{eq:SWfund1}) are related to $u$ and $m_{i}$
used in $\mathcal{Z}$ in a complicated way \cite{Dorey:1996bn,Argyres:1999ty},
\begin{equation}
\hat{u}=h_{u}\left(u,\mathtt{q},m_{i}\right),\quad\hat{m}_{i}=h_{i}\left(\mathtt{q},m_{i}\right).
\end{equation}
Due to this problem, it is complicated to compare our results directly with
this form of the Seiberg-Witten curve.

A more conceptual reason why the Seiberg-Witten curve (\ref{eq:SWfund1})
is not suitable for the instanton counting is the following. Since
the Seiberg-Witten curve (\ref{eq:SWfund1}) is obtained from a mass
deformation of (\ref{eq:SWfund0}), the parameters appearing in (\ref{eq:SWfund1})
are measured in the limit $\hat{m}_{i}\to0$, or equivalently $a\to\infty$,
and $\tau_{\mathrm{SW}}$ is defined as
\begin{equation}
\tau_{\mathrm{SW}}=\frac{1}{2\pi\mathrm{i}}\left.\frac{\partial^{2}\mathcal{F}}{\partial a^{2}}\right|_{\hat{m}_{i}=0}.
\end{equation}
On the other hand, the partition function $\mathcal{Z}$ is computed
as a series expansion in $\mathtt{q}$, and the convergence of the
series requires that $\mathtt{q}$ is small. All the parameters appearing
in $\mathcal{Z}$ are naturally measured in the limit $\mathtt{q}\to0$,
which is also the degenerate limit of the punctured sphere in the
class $\mathcal{S}$ construction. Two limits $\mathtt{q}\to0$ and
$a\to\infty$ are the same for asymptotically free theories, but in
general are different for superconformal theories. 

To make life simple, we would like to work with another Seiberg-Witten curve. 
Our choice of the Seiberg-Witten curve is constructed from the qq-characters
of the theory. The fundamental qq-character of the $\mathrm{SU}(2)$
gauge theory with $N_{f}\leq4$ fundamental hypermultiplets is given
by \cite{Nekrasov:2015wsu,Jeong:2019fgx}
\begin{equation}
\mathscr{X}(x)=\mathscr{Y}\left(x+2\varepsilon_{+}\right)+\mathtt{q}\mathscr{Y}(x)^{-1}\prod_{i=1}^{N_{f}}\left(x+m_{i}+\varepsilon_{+}\right),
\end{equation}
where the observable $\mathscr{Y}(x)$ is the quantum corrected characteristic
polynomial of $\phi$ in the $\Omega$-background,
\begin{equation}
\mathscr{Y}(x)=x^{2}\exp\left(-\sum_{n=1}^{\infty}\frac{1}{nx^{n}}\mathrm{Tr}\phi^{n}\right)=x^{2}-\frac{1}{2}\mathrm{Tr}\phi^{2}+\mathcal{O}\left(x^{-1}\right).
\end{equation}
Although the expectation value of $\mathscr{Y}(x)$ contains singularities
in $x$, $\mathscr{X}(x)$ satisfies the non-perturbative Dyson-Schwinger
equation \cite{Nekrasov:2015wsu},
\begin{equation}
\left\langle \mathscr{X}(x)\right\rangle =\mathscr{T}(x),
\end{equation}
where $\mathscr{T}(x)$ is a quadratic polynomial in the variable
$x$ and can be fixed by comparing the large $x$ expansions of both
sides. For example, for $N_{f}=4$ we have
\begin{eqnarray}
\mathscr{T}(x) & = & \left\langle \left(\mathscr{X}(x)\right)_{+}\right\rangle \nonumber \\
 & = & \left(x+\varepsilon_{1}+\varepsilon_{2}\right)^{2}-\tilde{u}+\mathtt{q}\left(x^{2}+\left(\sum_{i=1}^{4}m_{i}\right)x+\tilde{u}\right),
\end{eqnarray}
where $\left(\cdot\right)_{+}$ means the polynomial part of the Laurent
series, and $\tilde{u}$ is identified with $u$ up to an additive
constant,
\begin{equation}
\tilde{u}=u+\sum_{n=1}^{\infty}\mathtt{q}^{n}f_{n}\left(m_{i}\right).
\end{equation}
It is not difficult to work out $f_{n}$ explicitly. However, we should
notice that in the proof of the non-perturbative Dyson-Schwinger equation
we consider $\mathrm{U}(N)$ gauge theories, and the relation between
$\tilde{u}$ and $u$ will be modified when we restrict ourselves
to gauge group $\mathrm{SU}(N)$. The Seiberg-Witten curve is given
by taking the flat space limit $\varepsilon_{1},\varepsilon_{2}\to0$,
\begin{equation}
Y+\frac{\mathtt{q}\prod_{i=1}^{N_{f}}\left(x+m_{i}\right)}{Y}=T(x),
\end{equation}
where
\begin{equation}
Y=\left\langle \mathscr{Y}(x)\right\rangle ,\quad T(x)=\lim_{\varepsilon_{1},\varepsilon_{2}\to0}\mathscr{T}(x)=\left(1+\mathtt{q}\right)x^{2}+\mathtt{q}\left(\sum_{i=1}^{4}m_{i}\right)x-\left(1-\mathtt{q}\right)\tilde{u},
\end{equation}
and the canonical Seiberg-Witten differential is given by
\begin{equation}
\lambda=x\frac{dY}{Y}.
\end{equation}
It is convenient to perform a change of variables,
\begin{equation}
y=\frac{2Y-T(x)}{1-\mathtt{q}},
\end{equation}
so that the Seiberg-Witten curve becomes
\begin{equation}
y^{2}=\left(\frac{T(x)}{1-\mathtt{q}}\right)^{2}-\frac{4\mathtt{q}}{\left(1-\mathtt{q}\right)^{2}}\prod_{i=1}^{N_{f}}\left(x+m_{i}\right).\label{eq:SWfund2}
\end{equation}
The right hand side of (\ref{eq:SWfund2}) is now a monic polynomial
in $x$ of degree four. The curve (\ref{eq:SWfund2}) can be viewed
as a hybrid of the Seiberg-Witten curve (\ref{eq:SWfund1}) and the
class $S$ curve \cite{Gaiotto:2009we,Gaiotto:2009hg}. It describes
a torus rather than a punctured sphere, but the parameters are measured in the same way as those in the class $\mathcal{S}$ curve. The Seiberg-Witten differential
$\lambda$ is determined by
\begin{equation}
\frac{\partial\lambda}{\partial\tilde{u}}=\frac{1}{2\pi\mathrm{i}}\frac{dx}{y},
\end{equation}
whose period integral gives
\begin{equation}
\frac{\partial a}{\partial\tilde{u}}=\frac{1}{2\pi\mathrm{i}}\oint_{A}\frac{dx}{y}.\label{eq:dadu}
\end{equation}

Using the result reviewed in appendix \ref{app:Period}, we can write
down the exact result of the period integral (\ref{eq:dadu}) in terms
of the hypergeometric function. We then expand it as
\begin{eqnarray}
\frac{\partial a}{\partial\tilde{u}} & = & \frac{1}{\tilde{u}^{1/2}}\left(\frac{1}{2}-\frac{\mathtt{q}}{8}-\frac{7\mathtt{q}^{2}}{128}-\frac{17\mathtt{q}^{3}}{512}\right)\nonumber \\
 &  & +\frac{1}{\tilde{u}^{3/2}}\left(\left\llbracket m_{1}^{2}\right\rrbracket \left(\frac{1}{128}\mathtt{q}^{2}+\frac{5}{512}\mathtt{q}^{3}\right)-\left\llbracket m_{1}m_{2}\right\rrbracket \left(\frac{1}{8}\mathtt{q}+\frac{3}{32}\mathtt{q}^{2}+\frac{41}{512}\mathtt{q}^{3}\right)\right)\nonumber \\
 &  & +\frac{1}{\tilde{u}^{5/2}}\left(\left\llbracket m_{1}^{2}m_{2}^{2}\right\rrbracket \left(\frac{9}{128}\mathtt{q}^{2}+\frac{63}{512}\mathtt{q}^{3}\right)+\mathrm{Pf}m\left(\frac{3}{8}\mathtt{q}+\frac{3}{4}\mathtt{q}^{2}+\frac{531}{512}\mathtt{q}^{3}\right)\right.\nonumber \\
 &  & +\left.\left\llbracket m_{1}^{2}m_{2}m_{3}\right\rrbracket \left(\frac{3}{32}\mathtt{q}^{2}+\frac{81}{512}\mathtt{q}^{3}\right)-\left\llbracket m_{1}^{3}m_{2}\right\rrbracket \left(\frac{3}{512}\mathtt{q}^{3}\right)\right)+\mathcal{O}\left(\mathtt{q}^{4},\frac{m_i^6}{\tilde{u}^{7/2}}\right).\label{eq:aufund}
\end{eqnarray}
We integrate (\ref{eq:aufund}) over $\tilde{u}$ to get $a\left(\tilde{u}\right)$,
and then solve the inversion $\tilde{u}(a)$,
\begin{eqnarray}
\tilde{u} & = & a^{2}\left(1+\frac{\mathtt{q}}{2}+\frac{13\mathtt{q}^{2}}{32}+\frac{23\mathtt{q}^{3}}{64}\right)\nonumber \\
 &  & +\left(\left\llbracket m_{1}^{2}\right\rrbracket \left(\frac{1}{32}\mathtt{q}^{2}+\frac{3}{64}\mathtt{q}^{3}\right)-\frac{1}{2}\left\llbracket m_{1}m_{2}\right\rrbracket \left(\mathtt{q}+\mathtt{q}^{2}+\mathtt{q}^{3}\right)\right)\nonumber \\
 &  & +\frac{1}{a^{2}}\left(\left\llbracket m_{1}^{2}m_{2}^{2}\right\rrbracket \left(\frac{1}{32}\mathtt{q}^{2}+\frac{3}{64}\mathtt{q}^{3}\right)\right.\nonumber \\
 &  & \left.+\mathrm{Pf}m\left(\frac{1}{2}\mathtt{q}+\frac{1}{2}\mathtt{q}^{2}+\frac{33}{64}\mathtt{q}^{3}\right)\right)+\mathcal{O}\left(\mathtt{q}^{4},\frac{m_i^6}{a^{4}}\right),
\end{eqnarray}
which matches $u$ computed in (\ref{eq:ufund}) up to $a$-independent
terms. It is easy to compute
\begin{eqnarray}
\log\left(\frac{du}{da}\right) & = & \log\left(\frac{d\tilde{u}}{da}\right)\nonumber \\
 & = & \log\left(2a\right)+\frac{\mathtt{q}}{2}+\frac{9\mathtt{q}^{2}}{32}+\frac{19\mathtt{q}^{3}}{96}\nonumber \\
 &  & -\frac{1}{a^{4}}\left(\frac{1}{32}\left\llbracket m_{1}^{2}m_{2}^{2}\right\rrbracket \left(\mathtt{q}^{2}+\mathtt{q}^{3}\right)\right.\nonumber \\
 &  & \left.+\mathrm{Pf}m\left(\frac{1}{2}\mathtt{q}+\frac{1}{4}\mathtt{q}^{2}+\frac{3}{16}\mathtt{q}^{3}\right)\right)+\mathcal{O}\left(\mathtt{q}^{4},\frac{m_i^6}{a^{6}}\right).\label{eq:uafund}
\end{eqnarray}

There are $6$ singularities on the Coulomb moduli space for the $\mathrm{SU}(2)$
gauge theory with $N_{f}=4$ fundamental hypermultiplets. Unlike the
previous case, it is complicated to write down the explicit expressions
of the discriminant loci where we have extra massless BPS states.
What we can do is to compute the physical discriminant $\Delta$ from
the mathematical discriminant $\hat{\Delta}$ of the Seiberg-Witten
curve (\ref{eq:SWfund2}) by dividing the $\tilde{u}^{6}$ coefficient
of $\hat{\Delta}$,
\begin{equation}
\Delta=\frac{\hat{\Delta}}{\mathrm{Coeff}_{\tilde{u}^{6}}\left(\hat{\Delta}\right)}.
\end{equation}
Then $\Delta$ is indeed a monic polynomial in $\tilde{u}$ of degree $6$. We can compute
\begin{eqnarray}
\log\Delta & = & 12\log\left(a\right)+3\mathtt{q}+\frac{27\mathtt{q}^{2}}{16}+\frac{19\mathtt{q}^{3}}{16}\nonumber \\
 &  & +\frac{1}{a^{2}}\left(\left\llbracket m_{1}^{2}\right\rrbracket \left(-1+\frac{3}{32}\mathtt{q}^{2}+\frac{3}{32}\mathtt{q}^{3}\right)\right)\nonumber \\
 &  & +\frac{1}{a^{4}}\left(-\frac{3}{8}\left\llbracket m_{1}^{2}m_{2}^{2}\right\rrbracket \left(\mathtt{q}^{2}+\mathtt{q}^{3}\right)\right.\nonumber \\
 &  & \left.-\mathrm{Pf}m\left(\frac{1}{2}+3\mathtt{q}+\frac{3}{2}\mathtt{q}^{2}+\frac{3}{2}\mathtt{q}^{3}\right)\right)+\mathcal{O}\left(\mathtt{q}^{4},\frac{m_i^6}{a^{6}}\right).\label{eq:Deltafund}
\end{eqnarray}

By comparing (\ref{eq:uafund})(\ref{eq:Deltafund}) with the explicit
calculation in the $\Omega$-background (\ref{eq:Afund})(\ref{eq:Bfund}),
we find that

\begin{equation}
A=\Lambda^{-\frac{1}{2}}\left(\frac{du}{da}\right)^{\frac{1}{2}},\quad B=\sqrt{2}\Lambda^{-\frac{3}{2}}\Delta^{\frac{1}{8}}.
\end{equation}
We see that the explicit dependence on $\mathtt{q}$ disappears. Therefore,
we confirm (\ref{eq:AB}), and we find the unambiguous ratio
\begin{equation}
\frac{\beta}{\alpha}=\sqrt{2}\Lambda^{-1}.
\end{equation}
Similar to the previous case, it is easy to check that we cannot get
(\ref{eq:AB}) if we use $m_{f}^{\prime}$ rather than $m_{f}$ as
the mass parameters. The strange $\Lambda$ dependence of $\beta / \alpha$ can be again understood as a remnant of gravitational couplings of the weakly gauged $\mathrm{Spin}(8)$ flavor symmetry. Each fundamental hypermultiplet contributes a factor of $\Lambda^{-\frac{1}{4}}$ to $\beta / \alpha$, and we have four fundamental hypermultiplets.

It is straightforward to perform the same computation in asymptotically
free theories with $N_{f}\le3$ fundamental hypermultiplets. The Seiberg-Witten
curve can be constructed in the same way from the fundamental qq-character. The expansion
(\ref{eq:expansion}) matches (\ref{eq:AB}) for each case, and the
overall factors $\alpha$ and $\beta$ depend on $\Lambda$ but not
on masses.

\section{Perturbative analysis in $\mathrm{SU}(N)$ super-Yang-Mills theory
\label{sec:Pert}}

In this section, we will study $A$ and $B$ in the $\mathrm{SU}(N)$ super-Yang-Mills theory. In particular, we would like to check the prediction (\ref{eq:alphabetaN}). For the purpose of determining $\alpha$ and $\beta$, it is sufficient to neglect the complicated instanton contributions and use only the perturbative part of the partition function.

%After explicitly checking the prediction of (\ref{eq:AB}) and computing $\alpha$ and $\beta$ in $\mathrm{SU}(2)$ gauge theories, it is definitely interesting to go on and study more general examples. However, the computation soon becomes rather complicated and Mathematica can take a long time to run. Here we would like to perform a perturbative analysis of $A$ and $B$ in the $\mathrm{SU}(N)$ super-Yang-Mills theory, and check the prediction (\ref{eq:alphabetaN}).

The one-loop partition function is given by
\begin{equation}
\mathcal{Z}^{\mathrm{1-loop}}=\prod_{i<j}\exp\left[-\gamma_{\varepsilon_{1},\varepsilon_{2}}\left(a_{i}-a_{j};\Lambda\right)-\gamma_{\varepsilon_{1},\varepsilon_{2}}\left(a_{i}-a_{j}-2\varepsilon_{+};\Lambda\right)\right],\label{eq:ZSUN}
\end{equation}
with the constraint
\begin{equation}
\sum_{i=1}^{N}a_{i}=0.
\end{equation}
We can expand (\ref{eq:ZSUN}) around the flat space limit (\ref{eq:expansion})
using (\ref{eq:gammaexp}),
\begin{eqnarray}
\mathcal{F}^{\mathrm{1-loop}} & = & \sum_{i<j}\left(\left(a_{i}-a_{j}\right)^{2}\log\left(\frac{a_{i}-a_{j}}{\Lambda}\right)-\frac{3}{2}\left(a_{i}-a_{j}\right)^{2}\right),\nonumber \\
\mathcal{H}^{\mathrm{1-loop}} & = & 0,\nonumber \\
\log A^{\mathrm{1-loop}} & = & \frac{1}{2}\sum_{i<j}\log\left(\frac{a_{i}-a_{j}}{\Lambda}\right),\nonumber \\
\log B^{\mathrm{1-loop}} & = & \frac{1}{2}\sum_{i<j}\log\left(\frac{a_{i}-a_{j}}{\Lambda}\right).\label{eq:pert}
\end{eqnarray}

We take the Seiberg-Witten curve to be \cite{Klemm:1994qs,Argyres:1994xh}
\begin{equation}
y^{2}=\left(\left\langle \det\left(x-\phi\right)\right\rangle \right)^{2}-4\Lambda^{2N},
\end{equation}
with the Coulomb branch order parameters
\begin{equation}
u_{n}=\left\langle \frac{1}{n}\mathrm{Tr}\phi^{n}\right\rangle ,\quad n=2,\cdots,N.
\end{equation}
Ignoring the instanton corrections, the Seiberg-Witten curve degenerates
to $y^{2}=\left(\left\langle \det\left(x-\phi\right)\right\rangle \right)^{2}$,
and $u_{n}$ are simply given by the classical result,
\begin{equation}
u_{n}=\frac{1}{n}\sum_{i=1}^{N}a_{i}^{n}.
\end{equation}
We take $a_{2},\cdots,a_{N}$ as independent parameters. Then we have
\begin{equation}
\left(\frac{du_{i}}{da_{j}}\right)^{\mathrm{1-loop}}=a_{j}^{i-1}-a_{1}^{i-1},
\end{equation}
and consequently
\begin{eqnarray}
\det\left(\frac{du_{i}}{da_{j}}\right)^{\mathrm{1-loop}} & = & \begin{vmatrix}a_{2}-a_{1} & \cdots & a_{N}-a_{1}\\
\vdots & \ddots & \vdots\\
a_{2}^{N-1}-a_{1}^{N-1} & \cdots & a_{N}^{N-1}-a_{1}^{N-1}
\end{vmatrix}\nonumber \\
 & = & \begin{vmatrix}1 & 0 & \cdots & 0\\
a_{1} & a_{2}-a_{1} & \cdots & a_{N}-a_{1}\\
\vdots & \vdots & \ddots & \vdots\\
a_{1}^{N-1} & a_{2}^{N-1}-a_{1}^{N-1} & \cdots & a_{N}^{N-1}-a_{1}^{N-1}
\end{vmatrix}\nonumber \\
 & = & \begin{vmatrix}1 & 1 & \cdots & 1\\
a_{1} & a_{2} & \cdots & a_{N}\\
\vdots & \vdots & \ddots & \vdots\\
a_{1}^{N-1} & a_{2}^{N-1} & \cdots & a_{N}^{N-1}
\end{vmatrix}\nonumber \\
 & = & \prod_{i<j}\left(a_{i}-a_{j}\right).
\end{eqnarray}
Meanwhile, the perturbative discriminant is given by
\begin{equation}
\Delta^{\mathrm{1-loop}}=\left[\prod_{i<j}\left(a_{i}-a_{j}\right)^{2}\right]^{2}.
\end{equation}
By comparing with (\ref{eq:pert}), we reproduce the expression (\ref{eq:AB}),
\begin{equation}
A=\Lambda^{-\frac{N(N-1)}{4}}\det\left(\frac{du_{i}}{da_{j}}\right)^{\frac{1}{2}},\quad B=\Lambda^{-\frac{N(N-1)}{4}}\Delta^{\frac{1}{8}}.
\end{equation}
Hence, we obtain
\begin{equation}
\alpha=\beta=\Lambda^{-\frac{N(N-1)}{4}},
\end{equation}
which confirms the prediction (\ref{eq:alphabetaN}). This also matches
our result (\ref{eq:ABpure}) when $N=2$. The overall numerical constants
of $\beta$ are different due to the different normalization of the
discriminant $\Delta$.

\section{Discussions and outlook \label{sec:Discussion}}

In this paper, we use the partition function in the $\Omega$-background
to compute explicitly the low energy effective couplings $A$ and
$B$ to topological invariants of the background gravitational field
in four-dimensional $\mathcal{N}=2$ supersymmetric gauge theories
with gauge group $\mathrm{SU}(2)$. We also study the $\mathrm{SU}(N)$
super-Yang-Mills theory at the perturbative level. Our results confirm
the previous predictions. We also determine the ratio of the overall factors $\beta / \alpha$. 
For $\mathrm{SU}(2)$ theory with either an adjoint hypermultiplet or four fundamental hypermultiplets, we find that $\beta / \alpha$ is independent of $\tau_{\mathrm{UV}}$. Nevertheless, $K_{u}\alpha^{\chi}\beta^{\sigma}$ can still be a nontrivial function of $\tau_{\mathrm{UV}}$. It would be interesting to have a better understanding of this fact. Since $\beta / \alpha$ naturally shows up in the blowup formula \cite{Moore:1997pc,Losev:1997tp}, it may be useful to analyze carefully the behavior of the $u$-plane integral under blowups for superconformal theories.

There is no conceptual problem in extending our computation to any
other $\mathcal{N}=2$ theory whose partition function in the $\Omega$-background
can be calculated. Technically, our brute force expansion in $\mathtt{q}$
can be rather complicated. It would be very interesting to see whether one could directly obtain
the all-instanton results using methods of topological recursion
\cite{Eynard:2004mh,Eynard:2007kz}. A possible strategy is to use
the theory of qq-characters \cite{Nekrasov:2015wsu,Nekrasov:2016qym,Nekrasov:2016ydq,Nekrasov:2017gzb,Nekrasov:2017rqy},
and generalize the derivation presented in \cite{Jeong:2017mfh,Jeong:2019fgx}.
This will be discussed in the future.

We should also point out that $A$ and $B$ were exactly computed
for the $\mathrm{SU}(2)$ super-Yang-Mills theory \cite{Nakajima:2003pg,Nakajima:2003uh}
and the $\mathrm{SU}(2)$ gauge theory with one fundamental hypermultiplet
\cite{Gottsche:2010ig} using the partition function in the $\Omega$-background
of the blowup $\widehat{\mathbb{C}^{2}}$. This blowup approach is
also powerful enough to determine the contact terms in the $u$-plane
integral. We shall discuss the generalization of this approach to
other gauge theories in a separate paper. Unfortunately, this blowup
approach is not always useful for superconformal theories due to the
lack of an important vanishing theorem.

The supersymmetric localization method allows us to provide a contour
integral formula for the exact partition function of $\mathcal{N}=2$
supersymmetric gauge theories on compact toric four-manifolds \cite{Hama:2012bg,Rodriguez-Gomez:2014eza,Festuccia:2018rew},
generalizing the pioneering work of Pestun \cite{Pestun:2007rz}.
It was shown in \cite{Bershtein:2015xfa,Bershtein:2016mxz} that the
equivariant Donaldson invariants can be calculated by explicitly evaluating
the contour integral for $\mathrm{U}(2)$ super-Yang-Mills theory
on $\mathbb{CP}^{2}$. These equivariant Donaldson polynomials correctly
reproduce ordinary Donaldson invariants in the limit $\varepsilon_{1},\varepsilon_{2}\to0$.
It would be interesting to have a better understanding of these computations
from the point of view of the $u$-plane integral, and to perform similar
computations with hypermultiplets.

%%%%%%%%%%%%%%%%%%%%Acknowledgments%%%%%%%%%%%%%%%%%%%%

\acknowledgments
We would like to thank Sungbong Chun, Saebyeok Jeong, Nikita Nekrasov, Du Pei, Samson Shatashvili, and Peng Zhao for helpful discussions. JM is supported by Laureate Award 15175 of the Irish Research Council. GM and XZ are supported in part by the U.S. Department of Energy under Grant No. DE-SC0010008.

%%%%%%%%%%%%%%%%%%%%Appendix%%%%%%%%%%%%%%%%%%%%

\appendix

\section{Special function $\gamma_{\varepsilon_{1},\varepsilon_{2}}\left(x;\Lambda\right)$
\label{app:Gamma}}

The special function $\gamma_{\varepsilon_{1},\varepsilon_{2}}\left(x;\Lambda\right)$
is defined through the zeta function regularization,
\begin{equation}
\gamma_{\varepsilon_{1},\varepsilon_{2}}\left(x;\Lambda\right)=\left.\frac{d}{ds}\right|_{s=0}\frac{\Lambda^{s}}{\Gamma(s)}\int_{0}^{\infty}\frac{dt}{t}t^{s}\frac{e^{-xt}}{\left(e^{\varepsilon_{1}t}-1\right)\left(e^{\varepsilon_{2}t}-1\right)}.
\end{equation}
It is related to Barnes' double Gamma function $\Gamma_{2}\left(\left.x\right|\varepsilon_{1},\varepsilon_{2}\right)$
by
\begin{equation}
\gamma_{\varepsilon_{1},\varepsilon_{2}}\left(x;1\right)=\log\Gamma_{2}\left(\left.x+\varepsilon_{1}+\varepsilon_{2}\right|\varepsilon_{1},\varepsilon_{2}\right).
\end{equation}
Let us define $\left\{ c_{n},n\in\mathbb{N}\right\} $ by
\begin{equation}
\frac{1}{\left(e^{\varepsilon_{1}t}-1\right)\left(e^{\varepsilon_{2}t}-1\right)}=\sum_{n=0}^{\infty}\frac{c_{n}}{n!}t^{n-2},
\end{equation}
where
\begin{equation}
c_{0}=\frac{1}{\varepsilon_{1}\varepsilon_{2}},\quad c_{1}=-\frac{\varepsilon_{1}+\varepsilon_{2}}{2\varepsilon_{1}\varepsilon_{2}},\quad c_{2}=\frac{\varepsilon_{1}^{2}+3\varepsilon_{1}\varepsilon_{2}+\varepsilon_{2}^{2}}{6\varepsilon_{1}\varepsilon_{2}}.
\end{equation}
Then the expansion of $\gamma_{\varepsilon_{1},\varepsilon_{2}}\left(x;\Lambda\right)$
around the flat space limit $\varepsilon_{1},\varepsilon_{2}\to0$
can be computed using analytic continuation,
\begin{eqnarray}
\gamma_{\varepsilon_{1},\varepsilon_{2}}\left(x;\Lambda\right) & = & \left.\frac{d}{ds}\right|_{s=0}\frac{\Lambda^{s}}{\Gamma(s)}\int_{0}^{\infty}dt\sum_{n=0}^{\infty}\frac{c_{n}}{n!}t^{s+n-3}e^{-xt}\nonumber \\
 & = & \left.\frac{d}{ds}\right|_{s=0}\frac{\Lambda^{s}}{\Gamma(s)}\sum_{n=0}^{\infty}\frac{c_{n}}{n!}\Gamma(s+n-2)x^{2-s-n}\nonumber \\
 & = & \frac{1}{\varepsilon_{1}\varepsilon_{2}}\left(-\frac{1}{2}x^{2}\log\left(\frac{x}{\Lambda}\right)+\frac{3}{4}x^{2}\right)-\frac{\varepsilon_{1}+\varepsilon_{2}}{2\varepsilon_{1}\varepsilon_{2}}\left(x\log\left(\frac{x}{\Lambda}\right)-x\right)\nonumber \\
 &  & -\frac{\varepsilon_{1}^{2}+3\varepsilon_{1}\varepsilon_{2}+\varepsilon_{2}^{2}}{12\varepsilon_{1}\varepsilon_{2}}\log\left(\frac{x}{\Lambda}\right)+\sum_{n=2}^{\infty}\frac{c_{n}}{n(n-1)(n-2)}x^{2-n}.
\end{eqnarray}
In this paper, we need the expansions of the following two combinations
\begin{eqnarray}
 &  & \gamma_{\varepsilon_{1},\varepsilon_{2}}\left(x;\Lambda\right)+\gamma_{\varepsilon_{1},\varepsilon_{2}}\left(x-2\varepsilon_{+};\Lambda\right)\nonumber \\
 & = & \frac{1}{\varepsilon_{1}\varepsilon_{2}}\left(-x^{2}\log\left(\frac{x}{\Lambda}\right)+\frac{3}{2}x^{2}\right)-\frac{\varepsilon_{1}^{2}+3\varepsilon_{1}\varepsilon_{2}+\varepsilon_{2}^{2}}{6\varepsilon_{1}\varepsilon_{2}}\log\left(\frac{x}{\Lambda}\right)+\cdots,\nonumber \\
 &  & \gamma_{\varepsilon_{1},\varepsilon_{2}}\left(x-\varepsilon_{+};\Lambda\right)\nonumber \\
 & = & \frac{1}{\varepsilon_{1}\varepsilon_{2}}\left(-\frac{1}{2}x^{2}\log\left(\frac{x}{\Lambda}\right)+\frac{3}{4}x^{2}\right)+\frac{\varepsilon_{1}^{2}+\varepsilon_{2}^{2}}{24\varepsilon_{1}\varepsilon_{2}}\log\left(\frac{x}{\Lambda}\right)+\cdots.\label{eq:gammaexp}
\end{eqnarray}
Notice that there is no $\frac{\varepsilon_{1}+\varepsilon_{2}}{\varepsilon_{1}\varepsilon_{2}}$-term
in the expansion of both combinations.

\section{Period integrals on elliptic curves \label{app:Period}}

A general elliptic curve can be written as
\begin{equation}
y^{2}=x^{4}-c_{1}x^{3}+c_{2}x^{2}-c_{3}x+c_{4}=\left(x-r_{1}\right)\left(x-r_{2}\right)\left(x-r_{3}\right)\left(x-r_{4}\right),
\end{equation}
where
\begin{equation}
c_{n}=\sum_{1\leq i_{1}<\cdots<i_{n}\leq4}r_{i_{1}}\cdots r_{i_{n}}.
\end{equation}
We assume that $r_{1}<r_{2}<r_{3}<r_{4}$ are all real. The general
case can be obtained by analytic continuation. We define the A-cycle
and the B-cycle to enclose the cut $\left[r_{1},r_{2}\right]$ and
$\left[r_{2},r_{3}\right]$, respectively. The period integrals of
the holomorphic one-form are
\begin{equation}
\Pi_{\gamma}=\frac{1}{2\pi\mathrm{i}}\oint_{\gamma}\frac{dx}{y}=\frac{1}{2\pi\mathrm{i}}\oint_{\gamma}\frac{dx}{\sqrt{\left(x-r_{1}\right)\left(x-r_{2}\right)\left(x-r_{3}\right)\left(x-r_{4}\right)}},\quad\gamma=A,B.
\end{equation}
In this paper, we only need the period integral $\Pi_{A}$ over the
A-cycle. In order to compute the integral, we consider a useful variable
change
\begin{equation}
x=\frac{\left(r_{2}-r_{1}\right)r_{4}t+\left(r_{4}-r_{2}\right)r_{1}}{\left(r_{2}-r_{1}\right)t+\left(r_{4}-r_{2}\right)},
\end{equation}
so that $x=r_{1},r_{2},r_{3},r_{4}$ are mapped to $t=0,1,\frac{1}{\kappa},\infty$,
with
\begin{equation}
\kappa=\frac{\left(r_{1}-r_{2}\right)\left(r_{3}-r_{4}\right)}{\left(r_{1}-r_{3}\right)\left(r_{2}-r_{4}\right)}.\label{eq:zr}
\end{equation}
Then we have
\begin{eqnarray}
\Pi_{A} & = & \left[\left(r_{1}-r_{3}\right)\left(r_{2}-r_{4}\right)\right]^{-\frac{1}{2}}\frac{1}{\pi}\int_{0}^{1}\frac{dt}{\sqrt{t\left(1-t\right)\left(1-zt\right)}}\nonumber \\
 & = & \left[\left(r_{1}-r_{3}\right)\left(r_{2}-r_{4}\right)\right]^{-\frac{1}{2}}{_{2}}F_{1}\left(\frac{1}{2},\frac{1}{2},1,\kappa\right),\label{eq:PA1}
\end{eqnarray}
where we used the integral representation of hypergeometric function
\begin{equation}
{_{2}}F_{1}\left(\alpha,\beta,\gamma;z\right)=\frac{\Gamma(\gamma)}{\Gamma(\alpha)\Gamma(\beta)}\int_{0}^{1}dx\,x^{\beta-1}\left(1-x\right)^{\gamma-\beta-1}\left(1-xz\right)^{-\alpha}.
\end{equation}
We also define the discriminant $\Delta$ of the elliptic curve to
be
\begin{eqnarray}
\Delta & = & \prod_{i<j}\left(r_{i}-r_{j}\right)^{2}\nonumber \\
 & = & -27c_{4}^{2}c_{1}^{4}-4c_{3}^{3}c_{1}^{3}+18c_{2}c_{3}c_{4}c_{1}^{3}+c_{2}^{2}c_{3}^{2}c_{1}^{2}+144c_{2}c_{4}^{2}c_{1}^{2}\nonumber \\
 &  & -4c_{2}^{3}c_{4}c_{1}^{2}-6c_{3}^{2}c_{4}c_{1}^{2}+18c_{2}c_{3}^{3}c_{1}-192c_{3}c_{4}^{2}c_{1}-80c_{2}^{2}c_{3}c_{4}c_{1}\nonumber \\
 &  & -27c_{3}^{4}+256c_{4}^{3}-4c_{2}^{3}c_{3}^{2}-128c_{2}^{2}c_{4}^{2}+16c_{2}^{4}c_{4}+144c_{2}c_{3}^{2}c_{4}.
\end{eqnarray}

In general, the expression for the roots can be rather complicated.
Moreover, in the formula (\ref{eq:PA1}), the four roots $r_{1},r_{2},r_{3},r_{4}$
are not on equal footing. Using the quadratic transformation identity
of the hypergeometric function \cite{Bateman:100233},
\begin{equation}
{_{2}}F_{1}\left(\alpha,\beta,2\beta;z\right)=\left(1-z\right)^{-\frac{1}{2}\alpha}{_{2}}F_{1}\left(\frac{1}{2}\alpha,\beta-\frac{1}{2}\alpha,\beta+\frac{1}{2};-\frac{z^{2}}{4(1-z)}\right),
\end{equation}
with $\alpha=\beta=\frac{1}{2}$, we get
\begin{equation}
\Pi_{A}=\xi^{-\frac{1}{4}}{_{2}}F_{1}\left(\frac{1}{4},\frac{1}{4},1,\tilde{\kappa}\right),\label{eq:PA2}
\end{equation}
where
\begin{eqnarray}
\xi & = & \left(r_{1}-r_{3}\right)\left(r_{2}-r_{3}\right)\left(r_{1}-r_{4}\right)\left(r_{2}-r_{4}\right),\nonumber \\
\tilde{\kappa} & = & -\frac{\kappa^{2}}{4(1-\kappa)}=-\frac{\Delta}{4\xi^{3}}.
\end{eqnarray}
We see that the formula (\ref{eq:PA2}) is now symmetric in $r_{1},r_{2}$
and $r_{3},r_{4}$, but not in all of them. We can further apply the
cubic transformation identities of the hypergeometric function \cite{Bateman:100233},
\begin{equation}
{_{2}}F_{1}\left(3\alpha,\frac{1}{3}-\alpha,2\alpha+\frac{5}{6};z\right)=\left(1-4z\right)^{-3\alpha}{_{2}}F_{1}\left(\alpha,\alpha+\frac{1}{3},2\alpha+\frac{5}{6};\frac{27z}{\left(4z-1\right)^{3}}\right),
\end{equation}
with $\alpha=\frac{1}{12}$ to obtain
\begin{equation}
\Pi_{A}=\rho^{-\frac{1}{4}}{_{2}}F_{1}\left(\frac{1}{12},\frac{5}{12},1,\frac{27\Delta}{4\rho^{3}}\right),\label{eq:PA3}
\end{equation}
where
\begin{equation}
\rho=\xi+\frac{\Delta}{\xi^{2}}=\frac{1}{4}\left(\sum_{i<j}\left(r_{i}-r_{j}\right)^{2}\right)^{2}-\frac{3}{4}\sum_{i<j}\left(r_{i}-r_{j}\right)^{4}=c_{2}^{2}-3c_{1}c_{3}+12c_{4}.
\end{equation}
This formula makes all the roots completely symmetric. Furthermore,
we no longer need to solve the roots for a given elliptic curve in
order to obtain the period $\Pi_{A}$, thereby making the computation
much simpler.

\section{Modular forms and theta functions \label{app:Modular}}

\paragraph{Eisenstein series}

Let $\tau\in\mathbb{H}$ and $q=e^{2\pi\mathrm{i}\tau}$. The Eisenstein
series $E_{2k}$ is defined by
\begin{eqnarray}
E_{2k} & = & \frac{1}{2\zeta(2k)}\sum_{\substack{m,n\in\mathbb{Z}\\
(m,n)\neq(0,0)
}
}\frac{1}{\left(m+n\tau\right)^{2k}}\nonumber \\
 & = & 1+\frac{2}{\zeta\left(1-2k\right)}\sum_{n=1}^{\infty}\frac{n^{2k-1}q^{n}}{1-q^{n}}\nonumber \\
 & = & 1+\frac{2}{\zeta\left(1-2k\right)}\sum_{n=1}^{\infty}\sigma_{2k-1}(n)q^{n},
\end{eqnarray}
where $\sigma_{p}(n)$ is the divisor sum, the sum of the $p$th
powers of the divisors of $n$. The following explicit expansions
of the Eisenstein series are useful,
\begin{eqnarray}
E_{2} & = & 1-24\sum_{n=1}^{\infty}\sigma_{1}(n)q^{n}=1-24q-72q^{2}-96q^{3}-168q^{4}+\mathcal{O}\left(q^{5}\right),\nonumber \\
E_{4} & = & 1+240\sum_{n=1}^{\infty}\sigma_{3}(n)q^{n}=1+240q+2160q^{2}+6720q^{3}+17520q^{4}+\mathcal{O}\left(q^{5}\right),\nonumber \\
E_{6} & = & 1-504\sum_{n=1}^{\infty}\sigma_{5}(n)q^{n}=1-504q-16632q^{2}-122976q^{3}-532728q^{4}+\mathcal{O}\left(q^{5}\right).
\end{eqnarray}
The Eisenstein series $E_{2k}$ is a modular form of weight $2k$
under the $\mathrm{SL}(2,\mathbb{Z})$ modular transformation for
$k\geq2$,
\begin{equation}
E_{2k}\left(\frac{a\tau+b}{c\tau+d}\right)=\left(c\tau+d\right)^{2k}E_{2k}\left(\tau\right),\quad a,b,c,d\in\mathbb{Z},\quad ad-bc=1.
\end{equation}
The space of modular forms of $\mathrm{SL}(2,\mathbb{Z})$ forms a
ring that is generated by $E_{4}(\tau)$ and $E_{6}(\tau)$. For $k=1$,
$E_{2}$ is quasi-modular,
\begin{equation}
E_{2}\left(\frac{a\tau+b}{c\tau+d}\right)=\left(c\tau+d\right)^{2}E_{2}\left(\tau\right)+\frac{6}{\pi\mathrm{i}}c\left(c\tau+d\right).
\end{equation}
All quasi-modular forms can be expressed as polynomials of $E_{2}$,
$E_{4}$ and $E_{6}$. The derivatives of the Eisenstein series are
given by
\begin{eqnarray}
q\frac{dE_{2}}{dq} & = & \frac{E_{2}^{2}-E_{4}}{12},\nonumber \\
q\frac{dE_{4}}{dq} & = & \frac{E_{2}E_{4}-E_{6}}{3},\nonumber \\
q\frac{dE_{6}}{dq} & = & \frac{E_{2}E_{6}-E_{4}^{2}}{2}.\label{eq:dE}
\end{eqnarray}

\paragraph{Dedekind eta function}

The Dedekind eta function is defined by
\begin{equation}
\label{Deta}
\eta(\tau)=q^{\frac{1}{24}}\prod_{n=1}^{\infty}\left(1-q^{n}\right)=q^{\frac{1}{24}}\phi(q),
\end{equation}
where $\phi(q)$ is called the Euler function. Under the generators
of $\mathrm{SL}(2,\mathbb{Z})$, $\eta(\tau)$ transforms as
\begin{equation}
\eta(\tau+1)=e^{\frac{\pi\mathrm{i}}{12}}\eta(\tau),\quad\eta\left(-\frac{1}{\tau}\right)=\sqrt{-\mathrm{i}\tau}\eta(\tau).
\end{equation}
The derivative of $\eta(\tau)$ is related to $E_{2}$ by
\begin{equation}
q\frac{d}{dq}\log\eta(\tau)=\frac{E_{2}}{24}.
\end{equation}
We also use the expansion
\begin{equation}
\log\phi(q)=\sum_{n=1}^{\infty}\log\left(1-q^{n}\right)=-q-\frac{3}{2}q^{2}-\frac{4}{3}q^{3}-\frac{7}{4}q^{4}+\mathcal{O}\left(q^{5}\right).
\end{equation}

\paragraph{Jacobi theta functions}

The Jacobi theta functions are defined for two complex variables $z\in\mathbb{C}$
and $\tau\in\mathbb{H}$ as
\begin{eqnarray}
\label{Jthetas}
\theta_{1}\left(z;\tau\right) & = & \mathrm{i}\sum_{n\in\mathbb{Z}}\left(-1\right)^{n}w^{n+\frac{1}{2}}q^{\frac{1}{2}\left(n+\frac{1}{2}\right)^{2}},\nonumber \\
\theta_{2}\left(z;\tau\right) & = & \sum_{n\in\mathbb{Z}}w^{n+\frac{1}{2}}q^{\frac{1}{2}\left(n+\frac{1}{2}\right)^{2}},\nonumber \\
\theta_{3}\left(z;\tau\right) & = & \sum_{n\in\mathbb{Z}}w^{n}q^{\frac{1}{2}n^{2}},\nonumber \\
\theta_{4}\left(z;\tau\right) & = & \sum_{n\in\mathbb{Z}}\left(-1\right)^{n}w^{n}q^{\frac{1}{2}n^{2}},
\end{eqnarray}
where $w=e^{2\pi\mathrm{i}z}$ and $q=e^{2\pi\mathrm{i}\tau}$. When
evaluated at $z=0$, $\theta_{1}(0;\tau)=0$ and $\theta_{j}(\tau)=\theta_{j}(0;\tau)$
for $j=2,3,4$ satisfy
\begin{equation}
\theta_{2}(\tau)^{4}+\theta_{4}(\tau)^{4}=\theta_{3}(\tau)^{4},\quad\theta_{2}(\tau)\theta_{3}(\tau)\theta_{4}(\tau)=2\eta^{3}(\tau).
\end{equation}
They are also related to the Eisenstein series $E_{4}$ and $E_{6}$
by
\begin{equation}
E_{4}=\frac{1}{2}\left(\theta_{2}^{8}+\theta_{3}^{8}+\theta_{4}^{8}\right),\quad E_{6}=\frac{1}{2}\left(\theta_{2}^{4}+\theta_{3}^{4}\right)\left(\theta_{3}^{4}+\theta_{4}^{4}\right)\left(\theta_{4}^{4}-\theta_{2}^{4}\right),
\end{equation}
The transformation of $\theta_{j}(\tau)$ under the generators of
$\mathrm{SL}(2,\mathbb{Z})$ are
\begin{eqnarray}
\theta_{2}\left(-\frac{1}{\tau}\right) & = & \sqrt{-\mathrm{i}\tau}\theta_{4}(\tau),\quad\theta_{2}(\tau+1)=e^{\frac{\pi\mathrm{i}}{4}}\theta_{2}(\tau)\nonumber \\
\theta_{3}\left(-\frac{1}{\tau}\right) & = & \sqrt{-\mathrm{i}\tau}\theta_{3}(\tau),\quad\theta_{3}(\tau+1)=\theta_{4}(\tau),\nonumber \\
\theta_{4}\left(-\frac{1}{\tau}\right) & = & \sqrt{-\mathrm{i}\tau}\theta_{2}(\tau),\quad\theta_{4}(\tau+1)=\theta_{3}(\tau).
\end{eqnarray}

\section{Weierstrass's elliptic function \label{app:Weierstrass}}

Let $z$ be a coordinate of the torus, which can be viewed as the
complex plane with the identification $z\sim z+\pi\sim z+\pi\tau$.
We define Weierstrass's elliptic function $\wp\left(z;\tau\right)$
to be a meromorphic function in the complex plane with a double pole
at each lattice point,
\begin{eqnarray}
\wp\left(z;\tau\right) & = & \wp\left(z;\pi,\pi\tau\right)\nonumber \\
 & = & \frac{1}{z^{2}}+\sum_{\substack{m,n\in\mathbb{Z}\\
(m,n)\neq(0,0)
}
}\left[\frac{1}{\left(z+m\pi+n\pi\tau\right)^{2}}-\frac{1}{\left(m\pi+n\pi\tau\right)^{2}}\right],
\end{eqnarray}
satisfying the doubly periodic condition,
\begin{equation}
\wp\left(z;\tau\right)=\wp\left(z+\pi;\tau\right)=\wp\left(z+\pi\tau;\tau\right).
\end{equation}

The function $\wp\left(z;\tau\right)$ satisfies the differential
equation
\begin{equation}
\left(\wp^{\prime}\right)^{2}=4\wp^{3}-g_{2}\wp-g_{3}=4\left(\wp-e_{1}\right)\left(\wp-e_{2}\right)\left(\wp-e_{3}\right),\label{eq:wp}
\end{equation}
where
\begin{equation}
g_{2}=\frac{4}{3}E_{4}\left(\tau\right),\quad g_{3}=\frac{8}{27}E_{6}\left(\tau\right),
\end{equation}
and the roots $e_{1},e_{2},e_{3}$ can be expressed in terms of Jacobi
theta functions as
\begin{eqnarray}
e_{1} & = & \frac{1}{3}\left(\theta_{3}^{4}+\theta_{4}^{4}\right),\nonumber \\
e_{2} & = & -\frac{1}{3}\left(\theta_{2}^{4}+\theta_{3}^{4}\right),\nonumber \\
e_{3} & = & \frac{1}{3}\left(\theta_{2}^{4}-\theta_{4}^{4}\right).
\label{curveroots}
\end{eqnarray}
The modular discriminant $\Delta$ is defined as
\begin{equation}
\Delta=g_{2}^{3}-27g_{3}^{2}=\left(2\pi\right)^{12}\eta^{24}\left(\tau\right).
\end{equation}
When $\Delta>0$, all three are real and it is conventional to choose
$e_{1}>e_{2}>e_{3}$.

The function $\wp\left(z;\tau\right)$ is related to the Jacobi theta
function by
\begin{equation}
\wp\left(z;\tau\right)=-\frac{d^{2}}{dz^{2}}\log\theta_{1}\left(z;\tau\right)-\frac{1}{3}E_{2},
\end{equation}
where the constant term is fixed by comparing the Laurent expansion
of $\wp\left(z;\tau\right)$ at $z=0$,
\begin{equation}
\wp\left(z;\tau\right)=\frac{1}{z^{2}}+\frac{g_{2}}{20}z^{2}+\frac{g_{3}}{28}g_{3}z^{4}+\mathcal{O}\left(z^{6}\right),
\end{equation}
and the Laurent expansion of $\log\theta_{1}\left(z;\tau\right)$.

We are interested in calculating the integrals
\begin{equation}
\mathcal{P}_{n}=\frac{1}{\pi}\oint_{A}\wp^{n}dz,
\end{equation}
where we define the A-cycle to be $0\leq z\leq \pi$.
By definition, we have
\begin{eqnarray}
\mathcal{P}_{0} & = & \frac{1}{\pi}\oint_{A}dz=1,\\
\mathcal{P}_{1} & = & \frac{1}{\pi}\oint_{A}\wp dz=\frac{1}{\pi}\oint_{A}\left(-\frac{d^{2}}{dz^{2}}\log\theta_{1}\left(z;\tau\right)-\frac{1}{3}E_{2}\right)dz=-\frac{1}{3}E_{2}.
\end{eqnarray}
For $n=2$, we can obtain from the derivative of (\ref{eq:wp}) that
\begin{equation}
2\wp^{\prime\prime}=12\wp^{2}-g_{2},\label{eq:f2}
\end{equation}
from which we get
\begin{equation}
\mathcal{P}_{2}=\frac{1}{\pi}\oint_{A}\wp^{2}dz=\frac{1}{\pi}\oint_{A}\left(\frac{1}{6}\wp^{\prime\prime}+\frac{1}{12}g_{2}\right)dz=\frac{1}{12}g_{2}.
\end{equation}
The period integrals $\mathcal{P}_{n}$ for $n\geq3$ can be derived
recursively \cite{grosset2005elliptic}. In fact, using (\ref{eq:f2})
we have
\begin{eqnarray}
\mathcal{P}_{n} & = & \frac{1}{\pi}\oint_{A}\wp^{n}dz\nonumber \\
 & = & \frac{1}{\pi}\oint_{A}\wp^{n-2}\left(\frac{1}{6}\wp^{\prime\prime}+\frac{1}{12}g_{2}\right)dz\nonumber \\
 & = & \frac{1}{6\pi}\oint_{A}\wp^{n-2}\wp^{\prime\prime}dz+\frac{1}{12}g_{2}\mathcal{P}_{n-2}.
\end{eqnarray}
Integrating by parts the first term and substituting (\ref{eq:wp})
gives
\begin{eqnarray}
\oint_{A}\wp^{n-2}\wp^{\prime\prime}dz & = & -\left(n-2\right)\oint_{A}\wp^{n-3}\left(4\wp^{3}-g_{2}\wp-g_{3}\right)dz\nonumber \\
 & = & -\left(n-2\right)\oint_{A}\left(4\wp^{n}-g_{2}\wp^{n-2}-g_{3}\wp^{n-3}\right)dz.
\end{eqnarray}
Therefore, we obtain the following recurrence relation
\begin{equation}
\mathcal{P}_{n}=\frac{2n-3}{8n-4}g_{2}\mathcal{P}_{n-2}+\frac{n-2}{4n-2}g_{3}\mathcal{P}_{n-3},\quad n\geq3.
\end{equation}
Here we list the first few explicit expressions for $\mathcal{P}_{n}$,
$n\geq2$, as polynomials in $E_{2}$, $E_{4}$ and $E_{6}$,
\begin{eqnarray}
\mathcal{P}_{2} & = & \frac{1}{9}E_{4},\nonumber \\
\mathcal{P}_{3} & = & -\frac{1}{15}E_{2}E_{4}+\frac{4}{135}E_{6},\nonumber \\
\mathcal{P}_{4} & = & \frac{5}{189}E_{4}^{2}-\frac{8}{567}E_{2}E_{6},\nonumber \\
\mathcal{P}_{5} & = & -\frac{7}{405}E_{2}E_{4}^{2}+\frac{16}{1215}E_{4}E_{6}.
\end{eqnarray}

%%%%%%%%%%%%%%%%%%%%References%%%%%%%%%%%%%%%%%%%%

\bibliographystyle{JHEP}

\providecommand{\href}[2]{#2}\begingroup\raggedright\endgroup

\end{document}